%% file: florah.tex
\documentclass[fleqn,usenatbib]{mnras}

\usepackage{newtxtext,newtxmath}
\usepackage[T1]{fontenc}

\DeclareRobustCommand{\VAN}[3]{#2}
\let\VANthebibliography\thebibliography
\def\thebibliography{\DeclareRobustCommand{\VAN}[3]{##3}\VANthebibliography}

\usepackage{graphicx}	% Including figure files
\usepackage{amsmath}	% Advanced maths commands
\usepackage{mathtools}
\usepackage{fontawesome}
\usepackage{hyperref}
\usepackage{xspace}
\usepackage{dsfont}
\usepackage{graphicx}
\usepackage{graphbox}
\usepackage{siunitx}
\usepackage{bbm}
\usepackage{soul}
\usepackage{float}
\usepackage{tabularx}
\usepackage{multirow}
\usepackage{dcolumn}
\usepackage{makecell}
\usepackage{afterpage}
\usepackage{longtable}
\usepackage{adjustbox}
\usepackage[justification=centering]{caption}

\newcommand{\florah}{\textsc{florah}\xspace}
\newcommand{\florahv}{\textsc{florah-v}\xspace}
\newcommand{\florahg}{\textsc{florah-g}\xspace}
\newcommand{\florahvg}{\textsc{florah-vg}\xspace}
\newcommand{\gureft}{\textsc{gureft}\xspace}
\newcommand{\gureftc}{\textsc{gureft-c}\xspace}
\newcommand{\vsmdpl}{\textsc{vsmdpl}\xspace}

\newcommand{\Mvir}{\ensuremath{M_\mathrm{vir}}\xspace}
\newcommand{\logMvir}{\ensuremath{\log_{10} M_\mathrm{vir}}\xspace}
\newcommand{\cvir}{\ensuremath{c_\mathrm{vir}}\space}
\newcommand{\xt}[1]{\ensuremath{x^{({#1})}}}
\newcommand{\yt}[1]{\ensuremath{y^{({#1})}}}
\newcommand{\at}[1]{\ensuremath{a^{({#1})}}}
\newcommand{\Ht}[1]{\ensuremath{h^{({#1})}}}
\newcommand{\zt}[1]{\ensuremath{z^{({#1})}}}
\newcommand{\ct}[1]{\ensuremath{\cvir^{({#1})}}}
\newcommand{\Mt}[1]{\ensuremath{\Mvir^{({#1})}}}
\newcommand{\logMt}[1]{\ensuremath{\logMvir^{({#1})}}}

\title[Generative model for halo assembly histories]{FLORAH: A generative model for halo assembly histories}

\author[Nguyen et al.]{
Tri Nguyen,$^{1, 2, 3, 6}$\thanks{E-mail: tnguy@mit.edu}
Chirag Modi,$^{3, 4}$
L. Y. Aaron Yung,$^{5}$
and Rachel S. Somerville$^{3}$
\\
$^{1}$Kavli Institute for Astrophysics and Space Research, Massachusetts Institute of Technology, 77 Massachusetts Ave, Cambridge, MA 02139, USA\\
$^{2}$The NSF AI Institute for Artificial Intelligence and Fundamental Interactions, Cambridge, MA 02139, USA\\
$^{3}$Center for Computational Astrophysics, Flatiron Institute, 162 5th Avenue, New York, NY 10010, USA\\
$^{4}$Center for Computational Mathematics, Flatiron Institute, 162 5th Avenue, New York, NY 10010, USA\\
$^{5}$Astrophysics Science Division, NASA Goddard Space Flight Center, 8800 Greenbelt Rd, Greenbelt, MD 20771, USA\\
$^{6}$Department of Physics, Massachusetts Institute of Technology, 77 Massachusetts Ave, Cambridge, MA 02139, USA
}

\date{Accepted XXX. Received YYY; in original form ZZZ}

\pubyear{2023}

\begin{document}
\label{firstpage}
\pagerange{\pageref{firstpage}--\pageref{lastpage}}
\maketitle

\begin{abstract}
The mass assembly history (MAH) of dark matter halos plays a crucial role in shaping the formation and evolution of galaxies. 
MAHs are used extensively in semi-analytic and empirical models of galaxy formation, yet current analytic methods to generate them are inaccurate and unable to capture their relationship with the halo internal structure and large-scale environment.
This paper introduces \florah, a machine-learning framework for generating assembly histories of ensembles of dark matter halos. 
We train \florah on the assembly histories from the \gureft and \vsmdpl N-body simulations and demonstrate its ability to recover key properties such as the time evolution of mass and concentration.
We obtain similar results for the galaxy stellar mass versus halo mass relation and its residuals when we run the Santa Cruz semi-analytic model on \florah-generated assembly histories and halo formation histories extracted from an N-body simulation.
We further show that \florah also reproduces the dependence of clustering on properties other than mass (assembly bias), which is not captured by other analytic methods.
By combining multiple networks trained on a suite of simulations with different redshift ranges and mass resolutions, we are able to construct accurate main progenitor branches (MPBs) with a wide dynamic mass range from $z=0$ up to an ultra-high redshift $z \approx 20$, currently far beyond that of a single N-body simulation. 
\florah is the first step towards a machine learning-based framework for planting full merger trees; this will enable the exploration of different galaxy formation scenarios with great computational efficiency at unprecedented accuracy.
\end{abstract}


\section{Introduction}
\label{section:intro}

In the $\Lambda$-Cold Dark Matter ($\Lambda$CDM) paradigm, dark matter (DM) halos form hierarchically and grow in mass through the mergers of smaller DM halos~\citep{1978MNRAS.183..341W}.
The assembly history of a halo or galaxy refers to the sequence of events that led to its formation and growth over cosmic time.
It encompasses the accretion of matter, mergers with other halos or galaxies, and internal processes such as star formation and black hole growth.
In simulations, this is often represented in the form of ``merger trees'' i.e. progenitor and descendant halos that are linked across cosmic time.
In the modern paradigm of galaxy formation, the properties of galaxies (such as stellar mass and star formation rate) are believed to be closely linked to the assembly history of their halos and their formation environment.
Understanding this intricate halo-galaxy connection remains one of the key open questions of modern astrophysics.

N-body simulations provide a powerful tool to directly follow the formation and evolution of DM halos as they merge with other halos and interact with the large-scale environment~\citep{2020NatRP...2...42V}.
However, these simulations are computationally expensive, and their cost grows rapidly with the simulated volume and resolution.
Therefore, they are often run with only dark matter and without including baryonic physics.
It is not feasible to run a single simulation that can simultaneously capture the formation histories of halos from dwarf galaxies ($10^5 - 10^{10}~\si{M_\odot}$) to galaxy clusters ($10^{14} - 10^{15}~\si{M_\odot}$) up to high redshifts. 
It is currently challenging to run even a single dark matter-only simulation with a volume comparable to that of existing galaxy surveys and a mass resolution sufficient to accurately trace the merger histories of the halos hosting the galaxies that are detected in those surveys --- let alone next-generation surveys. 
Computational limitations also hinder the exploration of different structure formation scenarios and cosmological parameters. 

Semi-analytic models (SAMs) are commonly used to populate DM halo merger trees with galaxies.
SAMs apply simplified prescriptions for baryonic physics (e.g., radiative cooling, star formation, supernova, and AGN feedback prescriptions, etc.) within cosmological merger trees to track the formation and evolution of galaxies and forward model their observable properties~\citep{1994MNRAS.271..781C, 2000MNRAS.319..168C, somerville2015, 2019MNRAS.483.2983Y, 2021MNRAS.503.3698H, 2021MNRAS.506.4011E}.
SAMs require significantly fewer computational resources compared to N-body simulations and allow for exploration of a wider range of physical processes and parameters. 
They have been used to study many aspects of galaxy formation, ranging from large-scale clustering~(e.g.~\citealt{kauffmann1999, somerville2001, 2003ApJ...599...38B, 2021MNRAS.502.4858S,  Hadzhiyska2021, 2021MNRAS.505..492A, 10.1093/mnras/stac2297, 2022MNRAS.515.5416Y, 2023MNRAS.519.1578Y}) to galaxy properties (e.g., colors, metallicities, sizes, cold gas contents)~\citep{2012MNRAS.426.2142L, 2013MNRAS.436.1787L, somerville2015, Popping2017, 2019MNRAS.490.2855Y, 2020MNRAS.494.1002Y,  2020MNRAS.491.5795H, 2021MNRAS.503.4474Y} to supermassive black hole formation and active galactic nuclei feedback~\citep{somerville2008a, 2011MNRAS.410...53F, 2021MNRAS.508.2706Y} to reionization~\citep{2021MNRAS.503.3698H, 2021MNRAS.506..202U, 2021MNRAS.506..215H}.
Similarly, semi-empirical models such as \textsc{UniverseMachine} and \textsc{EMERGE} link galaxy observables to halo formation histories \citep{2019MNRAS.488.3143B, 2020MNRAS.499.5702B, 2018MNRAS.477.1822M}. 
Accounting for the dependence of halo clustering on properties other than mass (assembly bias) has been shown to be important for accurately interpreting galaxy clustering measurements \citep[e.g][]{2001astro.ph.11069W, Hadzhiyska2021, 10.1093/mnras/stac2297}. 
"Decorated" halo occupation distribution models have been proposed, which again require knowledge of the halo formation history and/or structural properties \citep[e.g.][]{2016MNRAS.460.2552H}. 

To overcome these computational challenges, previous works have developed analytic methods that rely on the Extended Press-Schechter (EPS) formalism~\citep{1991ApJ...379..440B, 1991MNRAS.248..332B}. 
EPS trees are constructed by sampling the conditional mass probability $p(M_1 | M_0, z_0, z_1)$ that a halo with mass $M_0$ at redshift $z_0$ had a mass of $M_1$ at an earlier redshift $z_1 > z_0$.
For halos of any given initial mass $M_0$ and redshift $z_0$, the algorithm uses Monte-Carlo methods and assumes the Markov property to construct its past merger histories~\citep[e.g.][]{1993MNRAS.262..627L, somerville_kolatt1999, zentner2007, 2008MNRAS.383..615N, 2008MNRAS.383..557P, 2015MNRAS.450.1514C, 2015MNRAS.452.1217C}.
However, this approach has many limitations. 
The resulting MAHs and merger trees are known to disagree with the results of N-body simulations at the factor of few levels \citep{2000MNRAS.316..479S, 2007MNRAS.379..689L, 2019MNRAS.485.5010B}.
Moreover, EPS-based techniques are fundamentally unable to capture the relationship between the assembly history, the halo structure, and the environment.
For a comparison between different EPS-based techniques for generating merger trees, we refer readers to \cite{2014MNRAS.440..193J}.

A recent study has applied Generative Adversarial Networks and Convolutional Neural Networks to generate ensembles of merger trees \citep{2022MNRAS.514.3692R}. 
Although this approach seems promising, this work did not incorporate the correlation between halo formation history and environment and did not show a complete set of diagnostic tests on the resulting merger tree ensembles. 
Another recent study using differential programming was able to accurately capture both the MAHs and assembly bias \citep{2021OJAp....4E...7H}. 
However, this approach smooths out the MAH of individual halos, making it difficult to capture merger events. 
It is also unclear how the halo structure and formation environment can potentially be included in the framework.

In this paper, we introduce \florah (FLOw-based Recurrent model for Assembly Histories) a deep generative model based on recurrent neural network and normalizing flows, to generate assembly histories of halos.
As a first step, we focus on generating the mass assembly histories (MAHs) and DM concentration histories of only the main progenitor branches (MPBs).
The MPB tracks the most massive progenitor of a halo and thus is the most important for understanding the assembly history.
MPBs can be naturally modeled as a time-ordered sequence; hence we use a recurrent neural network to learn their representative features.
Furthermore, for a halo of given mass and concentration, we are interested in learning the full distribution of possible assembly histories and hence we combine this recurrent network with a normalizing flow to learn this distribution.
Once trained, \florah can be used to generate the MAHs and DM concentration histories of MPBs consistent with N-body simulations. Although \florah only generates the MPBs of merger trees, it is the first framework to build in the correlation between halo structural properties and merger history.

To demonstrate this, we use the Santa Cruz SAM (SC-SAM)~\citep{somerville2015, Yung2023b}, with free parameters calibrated to reproduce a subset of observed galaxy properties at $z=0$ (see \citealt{10.1093/mnras/stac2297} for detailed calibration criteria), to populate \florah-generated MPBs with galaxies and show that \florah can correctly capture the assembly bias and environmental dependence of galaxies, which existing EPS-based methods cannot reproduce.
In addition, by combining multiple networks trained on different simulations with varying resolutions and redshift ranges, we are able to generate high-resolution MPBs to ultra-high redshifts, from $z=0$ to $z=20$.
This allows us to overcome the dynamic range limit of a single simulation.

This paper is organized as follows.
Section~\ref{section:simulation} provides a detailed description of the training simulation and the algorithms employed to extract the merger trees. 
Section~\ref{section:method} outlines our proposed method, including the data preprocessing procedure for a single simulation (Section~\ref{section:single_box}) and for the combination of multiple simulations (Section~\ref{section:combine_box}). 
We describe the neural network architecture and optimization in Section~\ref{section:forward_model}, followed by an explanation of the generation procedure in Section~\ref{section:generation}. 
In Section~\ref{section:result}, we show the generation results with \vsmdpl (Section~\ref{section:vsmdpl_result}) and \gureft (Section~\ref{section:gureft_result}), while Section~\ref{section:gureft_vsmdpl_result} presents a comparison with the combined \gureft and \vsmdpl simulations. 
We compare \florah to previous approaches for modeling MAHs, and discuss the limitations of the current approach and potential avenues for future research in Section~\ref{section:discussion}.
We conclude the paper in Section~\ref{section:conclusion}.

\section{Simulation}
\label{section:simulation}

\input{table_sim}

For training datasets, we extract merger trees from the \gureft (Gadget at Ultra-high Redshift with Extra Fine Timesteps)~\citep{gureft_paper}  suites and the \vsmdpl simulation from the MultiDark suite~\citep{multidark}.
Both \gureft and \vsmdpl are DM-only, N-body simulations run with the \textsc{Gadget-2} smoothed particle hydrodynamics code~\citep{gadget, gadget2}.
The cosmological parameters are broadly consistent with the Planck 2013 results~\citep{planck2013}, namely $\Omega_m = 0.307, \Omega_\Lambda = 0.693, h = 0.678, \sigma_8 = 0.823, n_s = 0.960$.
The specifications of \gureft and \vsmdpl are summarized in Table~\ref{tab:sim}.

The \gureft suite consists of four boxes with sizes of 5, 15, 35, and 90 $\si{Mpc \, h^{-1}}$. 
Each box has the same number of particles $1024^3$, with the smaller box having a higher mass resolution.
Therefore, each box captures the assembly histories of a different halo mass range.
The \gureft boxes have $171$ snapshots from $z \approx 40$ to $6$ at a high temporal resolution, making them the ideal simulations to capture the merger rates and assembly histories at high to ultra-high redshift. 
For $z \lesssim 6$, we instead use the \vsmdpl simulation, which has 151 snapshots from $z=0$ to $25$.
Thanks to its high mass resolution,  with a particle mass of $6.2 \times 10^6 \; \si{\mathrm{M}_\odot \, h^{-1}}$, \vsmdpl has a higher number of well-resolved and low-mass halos at high redshift.
This allows us to combine \vsmdpl with \gureft into one training dataset spanning a wider redshift and mass range.
We will describe this procedure in more detail in Section~\ref{section:combine_box}.

We construct merger trees for each \gureft simulation and use the publicly available merger trees from \vsmdpl.
All merger trees used in this work are constructed using the \textsc{Rockstar} halo finder and the \textsc{ConsistentTree} algorithm~\citep{2013ApJ...762..109B, 2013ApJ...763...18B}.
Note that merger trees may be sensitive to the details of the halo finding and tree construction algorithms. 
We plan to experiment with different parameters and tree reconstruction algorithms in future work. 

\begin{figure*}
    \centering
    \includegraphics[width=0.9\linewidth]{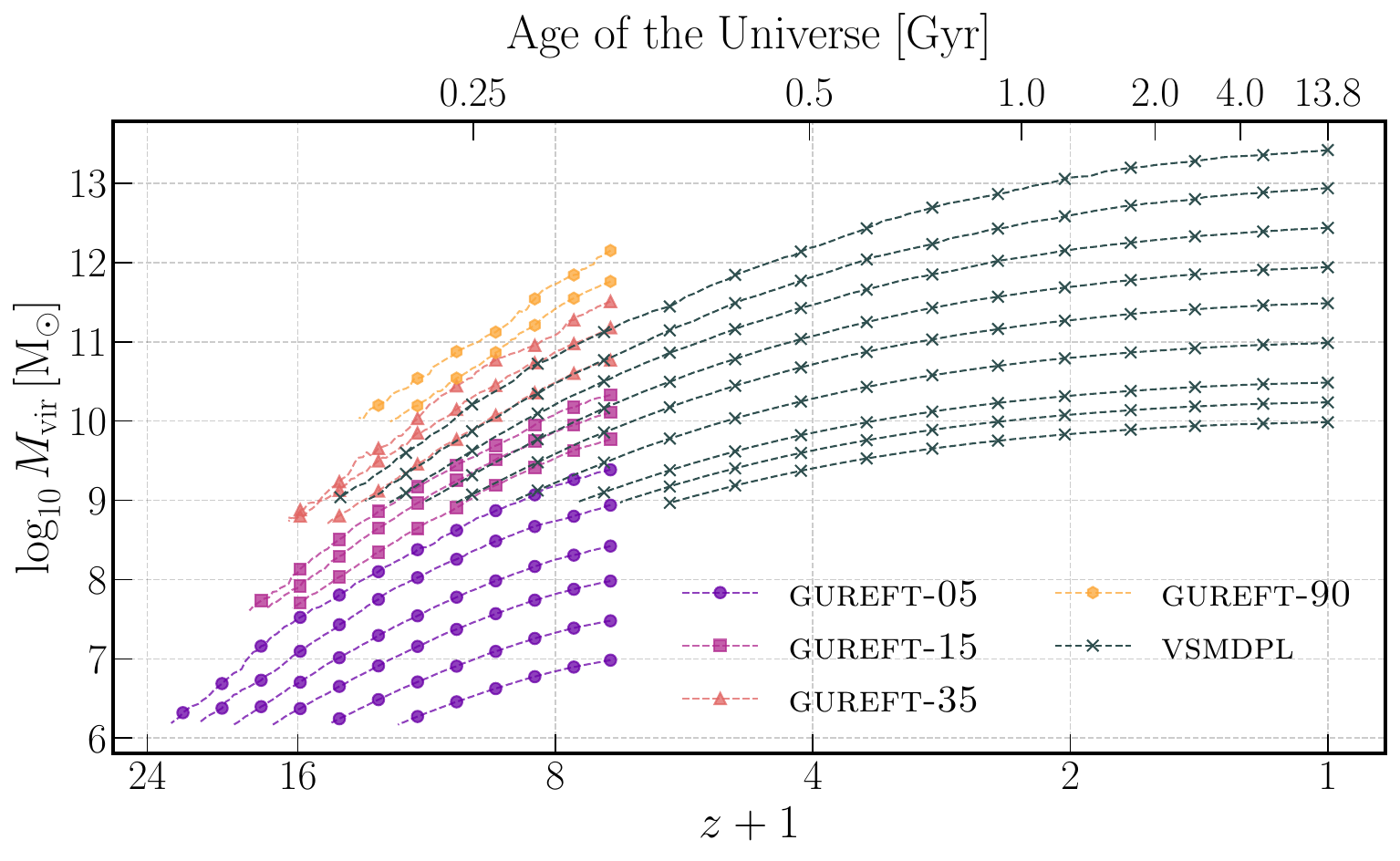}
    \caption{
    The median MAHs of the main progenitor branches of \gureft and \vsmdpl down to $100 \, M_\mathrm{DM}$, where $M_\mathrm{DM}$ is the mass of the DM particle in the corresponding simulations.
    }
    \label{fig:mean_mah_sim}
\end{figure*}

Figure~\ref{fig:mean_mah_sim} displays the median MAHs on the main branches of \gureft and \vsmdpl. The overlapping MAHs between the two simulations from $z \approx 13 - 5$ are crucial for combining them. 
We have only shown the median MAHs down to approximately $100 \, M_\mathrm{DM}$, where $M_\mathrm{DM}$ denotes the mass of the DM particles in the corresponding simulations. 
Halos with masses below this threshold cannot be reliably identified by \textsc{Rockstar}. 
Therefore, this limits the maximum redshift that we can generate for each simulation, with a maximum of $z \approx 20$ (from \gureft-05), as shown in Figure~\ref{fig:mean_mah_sim}.
Furthermore, the less massive root halos of \vsmdpl ($\lesssim 10^{10} \, \si{M_\odot}$) do not overlap with \gureft, which imposes a further limit on the maximum generation redshift for these halos. 
Additionally, each simulation has a unique particle mass and a resolved mass limit, which can complicate the process of combining the simulations. 
We will provide a more detailed discussion of the effects of particle mass and resolution limitations in Section~\ref{section:combine_box}.

\section{Methodology}
\label{section:method}

\begin{figure*}
    \centering
    \includegraphics[trim=-2cm 0 0 0cm, width=0.9\linewidth]{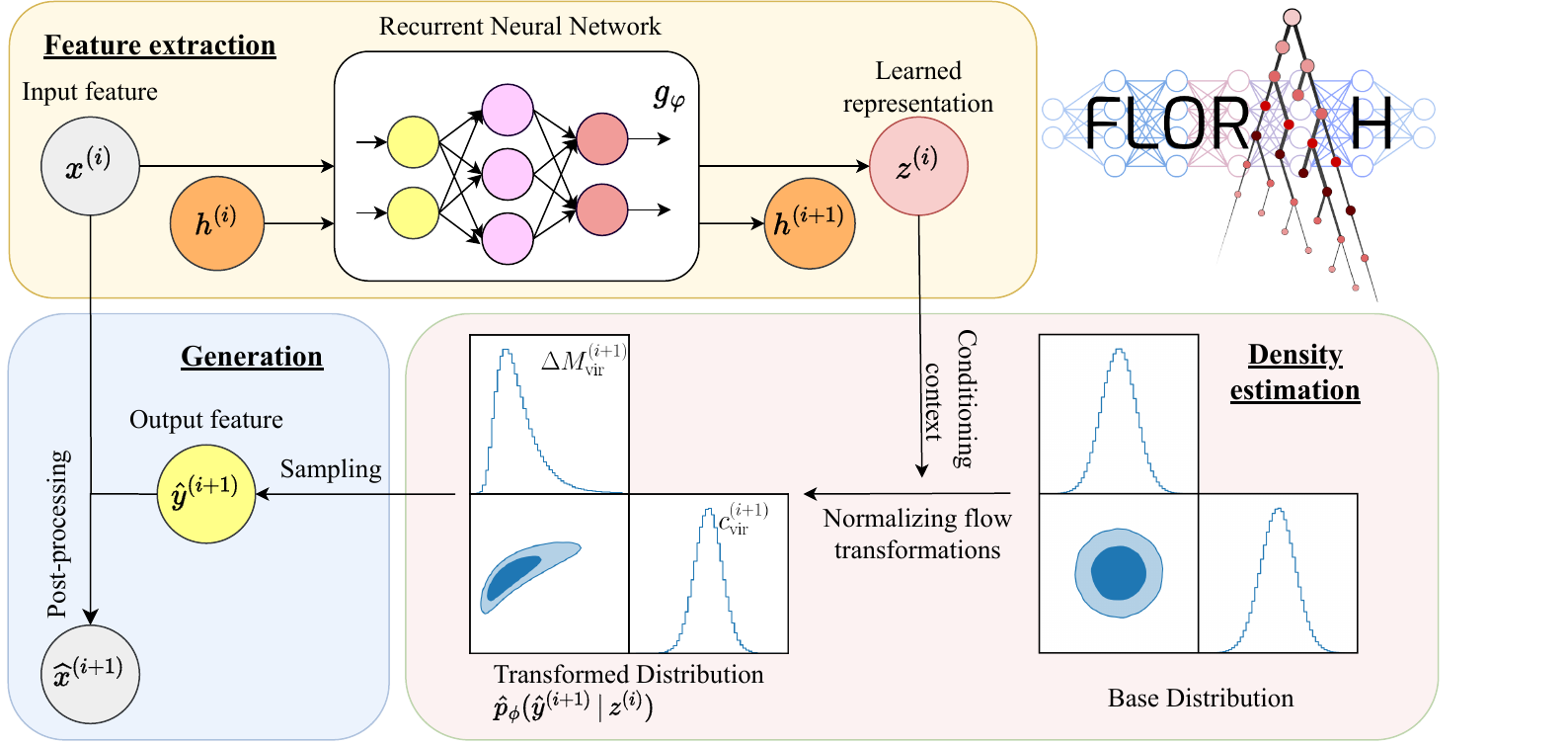}
    \caption{A schematic illustration of \florah.}
    \label{fig:pipeline}
\end{figure*}

In this section, we outline the data preprocessing, neural network architecture, and the optimization and generation process used in this work.
A schematic illustration of \florah is displayed in Figure~\ref{fig:pipeline}.
As a first approach, we aim to generate only the MPBs of merger trees since they contribute a first-order impact on the assembly histories of halos.
We plan to extend our framework to generate secondary branches and other branches in future work. 

\begin{figure}
    \centering
    \includegraphics[width=0.9\linewidth]{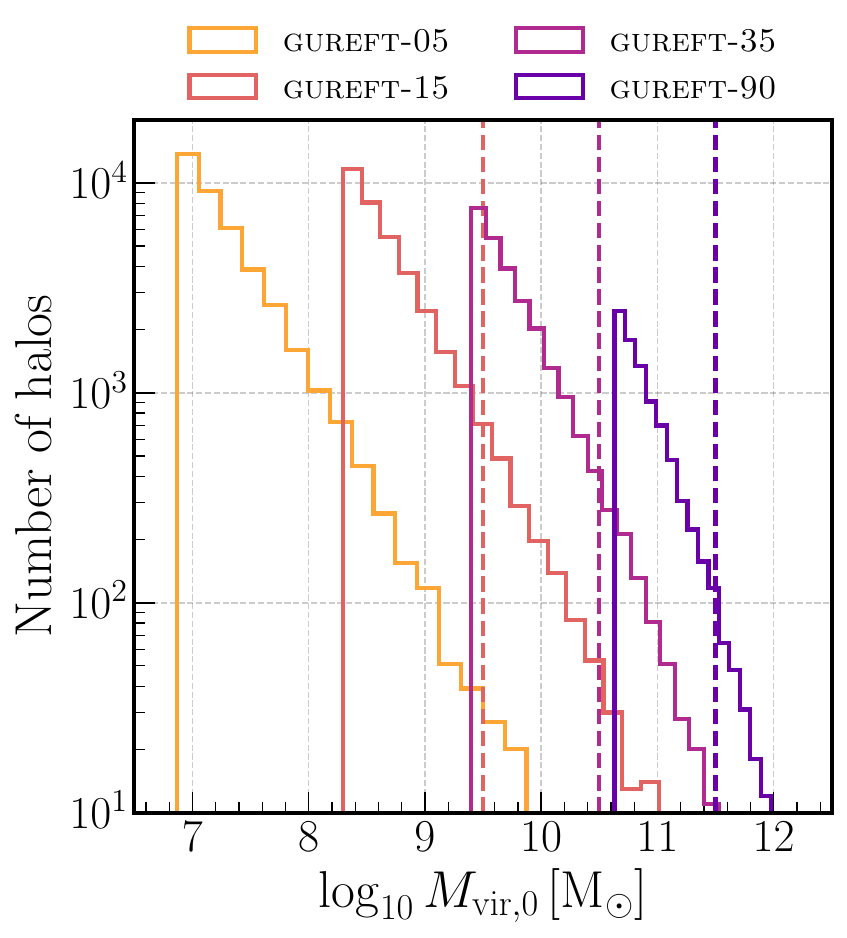}
    \caption{Distributions of \Mvir for four \gureft boxes at $z=5.89$. 
    Each dashed vertical line represents the additional minimum mass cut (on top of the 500 $M_\mathrm{DM}$ cut) in Section~\ref{section:combine_box} for the corresponding box in the combined box scheme. 
    We do not impose any additional minimum mass cut on \gureft-05.}
    \label{fig:gureft_mass_dist}
\end{figure}

\subsection{Data preprocessing}
\label{section:preprocessing}
 
\subsubsection{Using a single simulation}
\label{section:single_box}

In this section, we describe the data preprocessing steps for a single simulation.
The same steps apply for the four \gureft boxes and \vsmdpl with different parameters because they have different redshift ranges and resolutions.
We will highlight the parameters that differ in Section~\ref{section:result}, where we describe the training dataset in more detail. 
The following steps are the same for all simulations.

% \paragraph*{Extracting MPBs from simulations}
As a first step, we extract only the main progenitor branches (MPBs) and exclude all branches with root halos containing fewer than $500$ DM particles; halos below this limit may have poorly resolved progenitors. 
As previously mentioned in Section~\ref{section:simulation}, the lower limit of the progenitor mass is about $100 \, M_\mathrm{DM}$, below which \textsc{Rockstar} may not reliably identify halos.
We remove these unresolved halos during the post-processing of the generated MPBs (described in Section~\ref{section:generation}), rather than during the training dataset creation, as we have found that it leads to improved model performance. 
This resolution limit is particularly important when combining multiple simulations with varying $M_\mathrm{DM}$ into a single training dataset, which we further elaborate on in Section~\ref{section:combine_box}.
In addition, MPBs from \textsc{ConsistentTree} may sometimes have progenitor masses greater than their descendants. 
This is a common phenomenon that occurs due to the tree construction algorithm (including \textsc{ConsistentTree}) misidentifying the halos and their mass assignments. 
As with the unresolved halos, we will correct for this effect during the generation and post-processing of the MPBs in Section~\ref{section:generation}, rather than during the training phase.

% %\paragraph{Generalization by subsampling MPBs}
To enhance the generalization ability of our model, we employ data augmentation by creating multiple ``sub-branches'' for each MPB. 
We begin by selecting the initial snapshot from a uniform distribution up to the first 50 snapshots ($z \approx 2$ for \vsmdpl and $z \approx 10.6$ for \gureft). 
Next, we subsample the MPB by selecting every 2 to 6 snapshots (chosen uniformly) up to either a maximum redshift of $z_\mathrm{max, train}$ or a maximum length of 20 halos.
We limit the length of each sub-branch'' to 20 halos to prevent overly long sequences that could impact the learning process of the recurrent neural network~\citep{ pmlr-v28-pascanu13, pmlr-v108-ribeiro20a}.
Subsampling the tree in this way has a few advantages. 
First, it allows us to cover a wider range of redshifts, improving the overall generalization of the model. 
Second, it has a smoothing effect on the MAHs. 
Third, randomizing the time steps helps prevent the model from becoming overly reliant on one particular set of time steps and improves generalization.
We repeat the above data augmentation process 15 times for each tree.

% %\paragraph{Extracting features for the network:}
The input features to the recurrent neural network at each stage are the logarithm of the virial mass \logMvir, the Navarro-Frenk-White DM concentration \cvir~\citep{1996ApJ...462..563N}, and scale factors $a$ of the halo and the next progenitor.
The concentration is not given directly by \textsc{Rockstar} and instead computed from the relation $\cvir \equiv r_{\mathrm{vir}} / r_s$ where $r_s$ is the Navarro-Frenk-White (NFW) scale radius \citep{1996ApJ...462..563N} and $r_{\mathrm{vir}}$ is the virial radius.
Note that the NFW scale radius in \textsc{Rockstar} is computed in two different ways, by fitting the density profiles or by converting from the radius of the maximum circular velocity~\citep{2011ApJ...740..102K}.
Here we use the value provided by fit to the density profiles.
If a halo undergoes a recent major merger or has multiple density peaks, the fitting procedure in \textsc{Rockstar} can fail, and $r_s$ will be capped at $r_{\mathrm{vir}}$.This can create a ``pile up" in the concentration for halos with $r_s \approx r_{\mathrm{vir}}$.
Therefore, we require that $(r_s - r_{\mathrm{vir}}) / r_s > 0.1$, or equivalently $\cvir > 1.1$.
We remove all halos that do not pass this cut during the generation and post-processing of the MPBs in Section~\ref{section:generation}, as with the unresolved halos. 
Including \cvir helps the model learn assembly histories more accurately because \cvir has been found to capture the environmental dependency in many assembly bias studies~\citep{1999MNRAS.303..685N, 2002MNRAS.331...98V, 2006ApJ...652...71W, 2015MNRAS.450.1521C}.
We will experiment with expanding to more halo features (e.g. spin, shape) and environment features (e.g. local density, number of neighboring halos) in future work.
The scale factors of the halo and its next progenitor serve as the time features in our framework.
We have experimented with other time features like redshift and age and found similar performance.

For the output of the network, we model the logarithm of the \textit{change in} mass, defined as:
\begin{equation}
    \label{eq:accreted_mass}
    \Delta\logMt{i+1} = \log_{10} (\Mt{i+1} / \Mt{i}),
\end{equation} for the $(i+1)$-th halo, and the concentration \cvir of the next progenitor. 
We found that using accreted masses ($\Delta \logMt{i+1}$)  as targets, instead of progenitor masses ($\logMt{i+1}$), improves the model performance.
During generation, progenitor masses can be derived from masses and accreted masses using Eq.~\ref{eq:accreted_mass}.

\subsubsection{Combining multiple simulations}
\label{section:combine_box}

We present a procedure for combining the four \gureft boxes (\gureft-05, \gureft-15, \gureft-35, and \gureft-90) into a single training dataset, which we denote as \gureftc.
The four \gureft boxes exhibit significant overlap in halo masses as depicted by the distributions of \Mvir in Figure~\ref{fig:gureft_mass_dist}. 
However, due to differences in resolutions and volumes, MAH for halos in overlapping mass bins could vary by a significant amount, particularly at high redshifts. 
In such instances, the box with the highest resolution (i.e., the one with the lowest $M_\mathrm{DM}$) would yield a more accurate MAH and we would like to only use these for training. 
Hence to combine the \gureft boxes, we impose an additional minimum root mass cut of $\log_{10} M_\mathrm{min, root} \approx 9.5, \, 10.5, \, 11.5 \; \si{dex}$ for \gureft-15, \gureft-35, and \gureft-90 respectively. 
These cuts are applied directly to the root halos at $z=5.89$ and before any data augmentation step. 
These thresholds are chosen to strike a good balance between MAH accuracy (which increases for smaller boxes at a given mass) and the number of halos in a mass bin (which is lower for smaller boxes), as having too few halos can adversely affect the training process. 
No additional cut is applied to \gureft-05, which has the highest resolution, resulting in an over-representation of low-mass halos in our training dataset. 
Despite the imbalance in the number of low and high-mass halos, we show in Section~\ref{section:result} that we are able to capture the high-mass halos accurately.
The remaining data preprocessing steps follow the guidelines outlined in Section~\ref{section:single_box}. 
Each \gureft box is preprocessed separately, and the resulting datasets are combined into a single, large training dataset denoted as \gureftc.

\subsection{Neural network architecture and optimization}
\label{section:forward_model}

We model each sub-branch as a sequence of $N$ halos, with the root halo denoted with the index zero. The input and target feature vectors are:
\begin{align}
    \vec{x} &= \{\xt{i} \in \mathbb{R}^{f_\mathrm{in}}\}  = \{\logMt{i}, \ct{i}, \at{i}, \at{i+1}\},\\
    \vec{y} &= \{\yt{i+1} \in \mathbb{R}^{f_\mathrm{out}}\}
    = \{\Delta\logMt{i+1}, \ct{i+1}\},
\end{align}
respectively, where $f_\mathrm{in}=4$, $f_\mathrm{out}=2$, and  $i=0,..., N-2$.
We do not include any ``end-of-sequence'' token in our framework, so the feature vector includes only the first $N-1$ halos.
During the generation process (Section~\ref{section:generation}), we can choose to terminate the assembly history at a maximum redshift or minimum progenitor mass.
Our goal is to learn the true conditional distribution of $\yt{i+1}$, denoted as $p(\yt{i+1} | \{\xt{\leq i}\})$.

%\paragraph{Recurrent network architecture}
We use a recurrent neural network $g_\varphi: \mathbb{R}^{N_\mathrm{in}} \rightarrow \mathbb{R}^H$ with parameters $\varphi$ to extract $H$ summary features from the input features of each halo. 
The summary features are then:
\begin{align}
    &\vec{z} = \{\zt{i} \in \mathbb{R}^H\} = \{g_\varphi(\xt{i}, \Ht{i})\}, \\
    &\Ht{i} = \zt{i-1} \; \text{if } i > 0 \; \text{else } 0.
\end{align}
The hidden state $\Ht{i}$ is dependent on the input features of all the previous halos $\xt{\leq i}$, which allows the network to ``memorize" the entire assembly history.
Our recurrent network consists of 4 Gated Recurrent Unit (GRU) layers, each with $H=128$ hidden channels. 
For the recurrent layers, we have also experimented with different network architectures like Transformers~\citep{vaswani2023attention}, adding a decay mechanism to the hidden states~\citep{che2018recurrent}, and adding time-embedding layers.
We found that these choices result in similar performance while being more computationally intensive than our current model. 

%\paragraph{Normalizing flow architecture}
To learn the true probability distribution $p(\yt{i+1} \, | \, \xt{\leq i})$, we use a normalizing flow~\citep{DBLP:conf/icml/RezendeM15, 10.5555/3294771.3294994, papamakarios2019normalizing} that is conditioned on the summary features $\zt{i}$.
The flow thus estimates a conditional probability distribution $\hat{p}_\phi(\yt{i+1} \, | \, g_\varphi (\xt{i}, \Ht{i}))$ with learnable parameters $\phi$.
Our normalizing flow model consists of 4 Masked Autoregressive Flow (MAF) transformations. 
Each MAF includes a 4-layer Masked Autoencoder for Distribution Estimation (MADE) with a hidden dimension of 128~\citep{DBLP:conf/icml/GermainGML15,10.5555/3294771.3294994}.

%\paragraph{Training}
During training, we optimize the parameters $\{\varphi, \phi\}$ of the recurrent network and flow simultaneously using the negative log-density
\begin{equation}
    \label{eq:loss}
    \mathcal{L} = -\sum_{i=0}^{N-1} \log \hat{p}_\phi(\yt{i+1} \, | \, g_\varphi (\xt{i}, \Ht{i})),
\end{equation}
as the optimization loss. 
We use the AdamW optimizer~\citep{kingma2014adam,adamw2019} with parameters $(\gamma, \beta_1, \beta_2, \lambda) = (0.001, 0.9, 0.98, 0.01)$, where $\gamma$ is the learning rate, $\beta_1$ and $\beta_2$ are the running average coefficients, and $\lambda$ is the weight decay coefficient. 
At the end of each \textit{epoch} (defined as one full iteration over the training set), we evaluate the loss on the validation samples and reduce the learning rate by a factor of $10$ if no improvement is seen after 20 epochs.
We terminate the training process if the validation has not improved after 40 epochs.
The training process terminated at $\sim 200$ epochs or $\sim 2$ hours on a single NVIDIA Tesla V100 GPU.

\subsection{Generation and post-processing}
\label{section:generation}

\begin{figure*}
    \centering
    \includegraphics[width=0.9\textwidth]{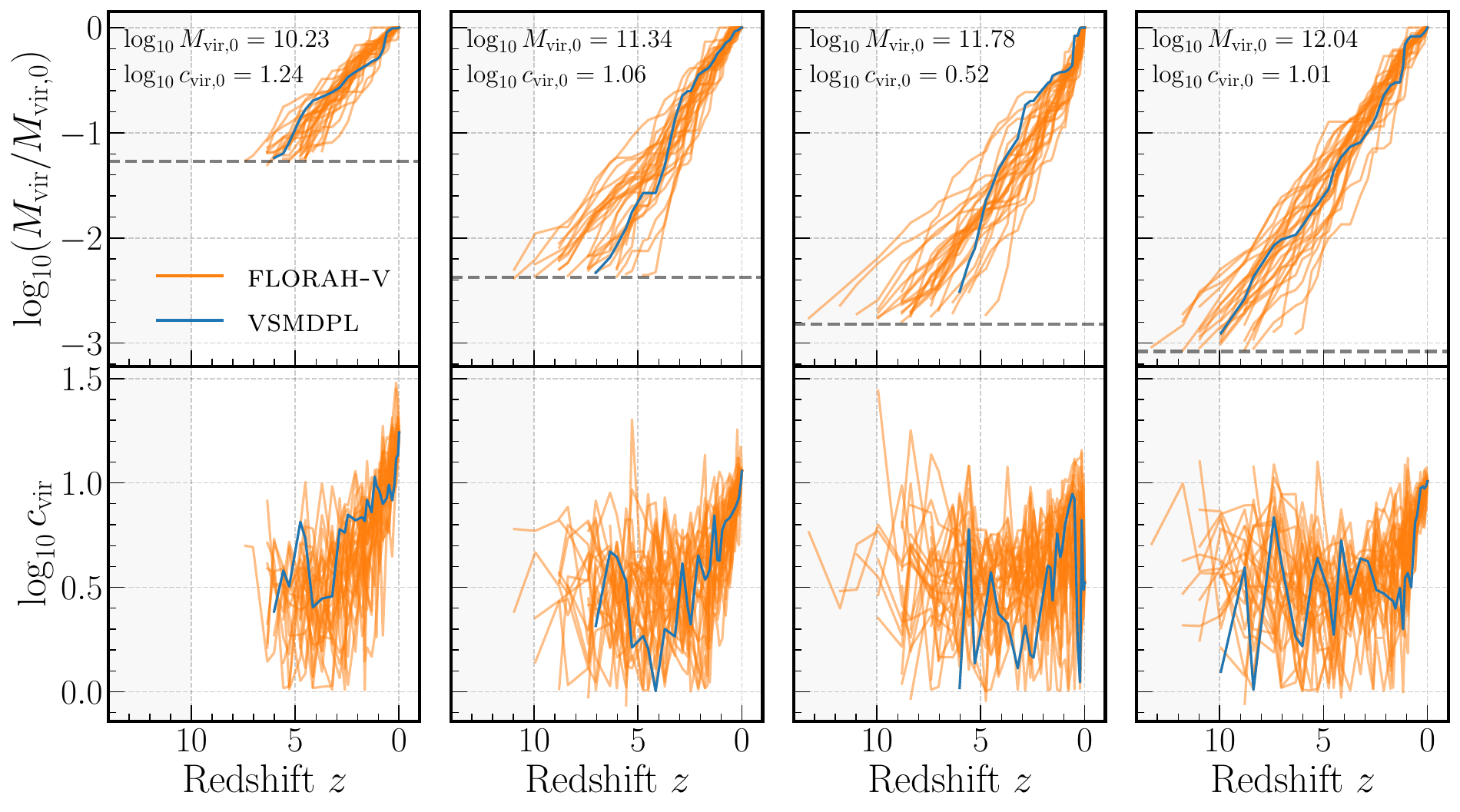}
    \caption{
    Example mass and concentration assembly histories generated by \florahv.
    Each column shows the mass (top) and concentration (bottom) assembly histories of an example root halo from the \vsmdpl simulation (blue) and 30 different realizations by \florahv. 
    The shaded gray box denotes the ``extrapolation region'' as we only train \florahv up to a maximum training redshift of  $z=10$.
    }
    \label{fig:vsmdpl_example_all}
\end{figure*}

\begin{figure*}
    \centering
    \includegraphics[width=0.9\textwidth]{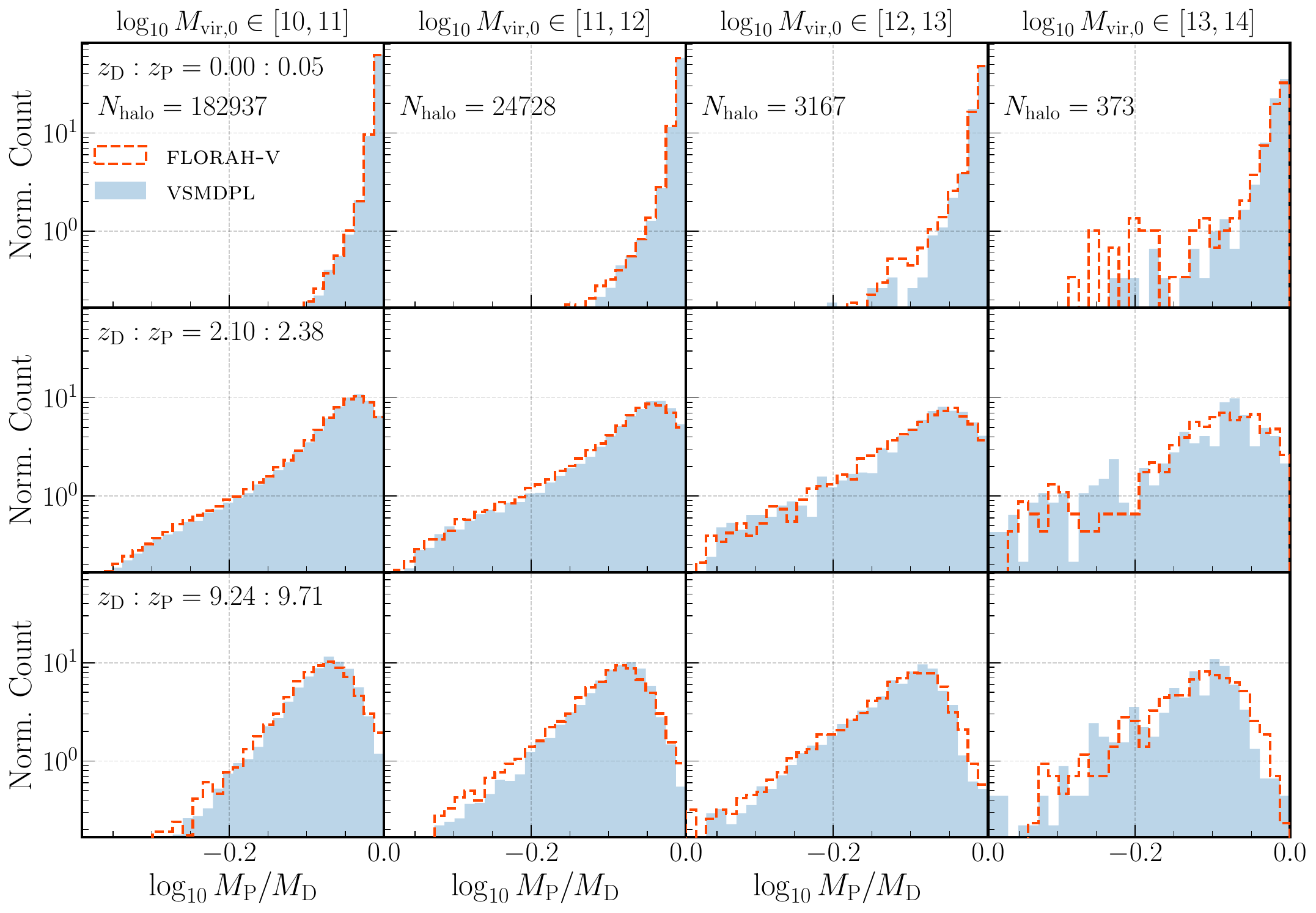}
    \caption{
    The normalized distribution of the progenitor-descendant mass ratio $\log_{10} M_P/M_D$ as modeled by \florahv (dashed) and from the \vsmdpl simulation (shaded).
    Each column represents a different root mass bin (in \si{M_\odot} unit).
    Each row represents a different progenitor-descendant redshift slice. 
    }
    \label{fig:vsmdpl_dlogm}
\end{figure*}

\begin{figure*}
    \centering
    \includegraphics[width=\linewidth]{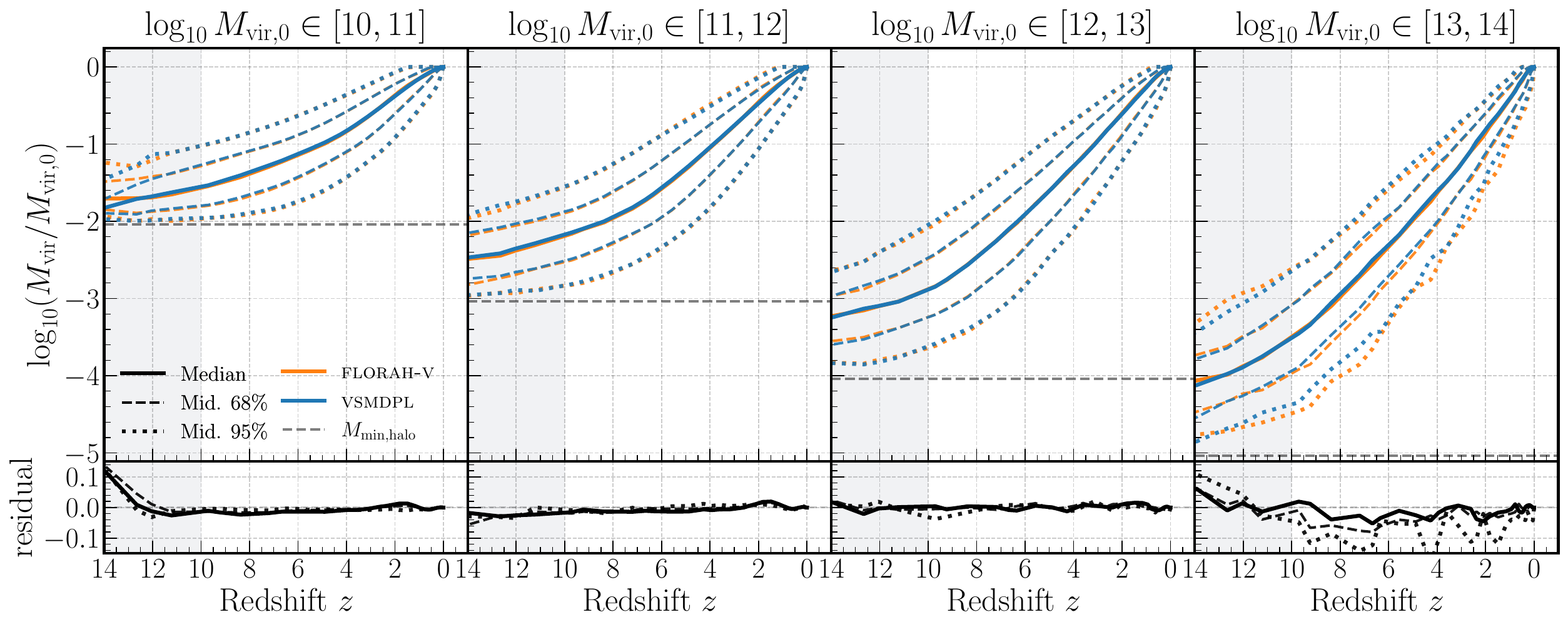}
    \includegraphics[width=\linewidth]{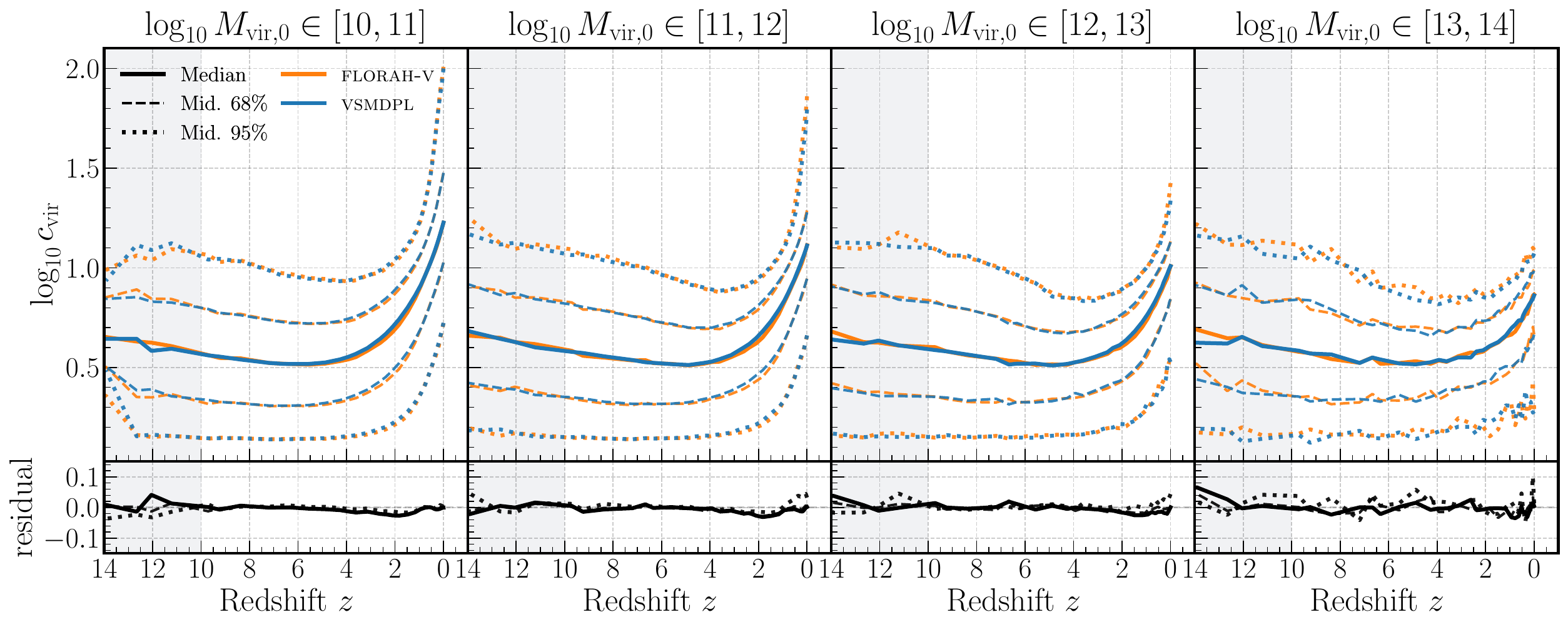}  
    \caption{
    The medians, middle 68-\% percentile, and middle-95\% containment regions of the MAHs (top) and DM concentration histories (bottom) of \vsmdpl (blue) and \florah (orange), along with their residuals (bottom), in four mass bins (in \si{M_\odot} unit). 
    The residual of the containment region is computed by averaging the residuals of the corresponding upper and lower percentile curves.
    The shaded gray box ($z > 10$) denotes the ``extrapolation region'' beyond the maximum training redshift.
    \label{fig:vsmdpl_mah_cah}
    }
\end{figure*}

Once the model is trained, the generation process for one tree is as follows:
\begin{enumerate}
    \item Select an initial halo feature \xt{0}, which is the initial mass $\log_{10} \Mt{0}$ and concentration $\ct{0}$.
    
    \item Select a list of scale factors $\{\at{i}\}$ as time steps. 
    \florah is robust across various time steps, as long as the chosen time steps remain within a reasonable range relative to the training time steps.

    \item Starting with the first time step $i=0$, pass the halo feature \xt{i} and the hidden state \Ht{i} (with \Ht{0} initialized to 0) through the GRU layers and extract the summary features $\zt{i} = g_\varphi(\xt{i}, \Ht{i})$.

    \item Sample $\hat{y}^{(i+1)} \sim \hat{p}_\phi(\hat{y}^{(i+1)} \, | \, \zt{i})$ using the normalizing flow layers.
    This returns the accreted mass $\Delta\logMt{i+1}$ and the DM concentration of the progenitor halo $\ct{i+1}$\footnote{
    As mentioned in Section~\ref{section:single_box}, \textsc{Rockstar} caps $r_s$ at $r_{\mathrm{vir}}$, resulting in a ``pile up'' at around $c \approx 1$. 
    To remove this population, we require the generated concentration $\ct{i+1} > 1.1$ and simply re-sample the halo if this condition is not satisfied.}.

    \item Convert the accreted mass to progenitor mass using Eq.~\ref{eq:accreted_mass} and set the feature of the progenitor halo to be $\xt{i+1} = (\logMt{i+1}, \ct{i+1}, \at{i+1}, \at{i+2})$.  
    
    \item Update the time step to $i+1$ and repeat Steps (ii) to (v) until a minimum progenitor mass $M_\mathrm{min, halo}$ or a maximum redshift $z_\mathrm{max, gen}$.
    We elaborate on the procedure to choose these thresholds below.
\end{enumerate}

As mentioned in Section~\ref{section:preprocessing}, halos with fewer than 100 DM particles are not reliably identified by \textsc{Rockstar}.
For a model trained on a single simulation (i.e. \vsmdpl in our case), we set the minimum progenitor mass to be $M_\mathrm{min, halo} = 100 \, M_\mathrm{DM}$, where $M_\mathrm{DM}$ is the DM particle mass of the simulation. 
For a model trained on a combined simulation (i.e. \gureftc in our case), due to the additional mass cut described in Section~\ref{section:combine_box}, the generated MPBs will have varying mass resolutions depending on their root masses. 
In the case of \gureftc, we determine the DM particle mass $M_\mathrm{DM}$ based on the root mass of the MPB at $z=5.89$ and set $M_\mathrm{min, halo} = 100 \, M_\mathrm{DM}$ as before.
The minimum progenitor masses are shown in the last column of Table~\ref{tab:sim} for each simulation.
As for the maximum generation redshift $z_\mathrm{max, gen}$, we generally recommend $z_\mathrm{max, gen}$ to be smaller than $z_\mathrm{max, train}$ to avoid extrapolation. 
However, we will show in Section~\ref{section:result} that \florah can extrapolate the MPBs beyond $z_\mathrm{max, train}$.

%\paragraph{Montonicity of MPBs}
Throughout the formation process of an MPB, the mass of the halo increases monotonically.
However, as mentioned in Section~\ref{section:preprocessing}, due to the misidentification of halos and mass assignments, progenitor halos may sometimes be assigned a mass greater than that of their descendant halos.
For \florah-generated MPBs, this can be corrected by simply re-sampling the mass and concentration in Step (iv). 
However, for MPBs in N-body simulations, the specific corrective measures will depend on the nature of the problem and the characteristics of the simulation data.
Thus, to compare \florah-generated MPBs with N-body MPBs, we adopt the following simple method.
For an MPB with $N$ halos, starting from the last progenitor halo at time step $i=N-1$ (with the root halo at time step $0$), we compare its mass $\Mt{i}$ with the mass of the next halo $\Mt{i-1}$.
If $\Mt{i} > \Mt{i-1}$, we assume that no accretion occurs and set the mass and concentration of the next halo $(\Mt{i-1}, \ct{i-1})$ to be the progenitor's, i.e. $(\Mt{i}, \ct{i})$.
We then move forward to the next time step $i-1$ until we reach the root halo at time step $0$. 

\section{Results}
\label{section:result}

\begin{figure*}
    \centering
    \includegraphics[width=0.495\linewidth]{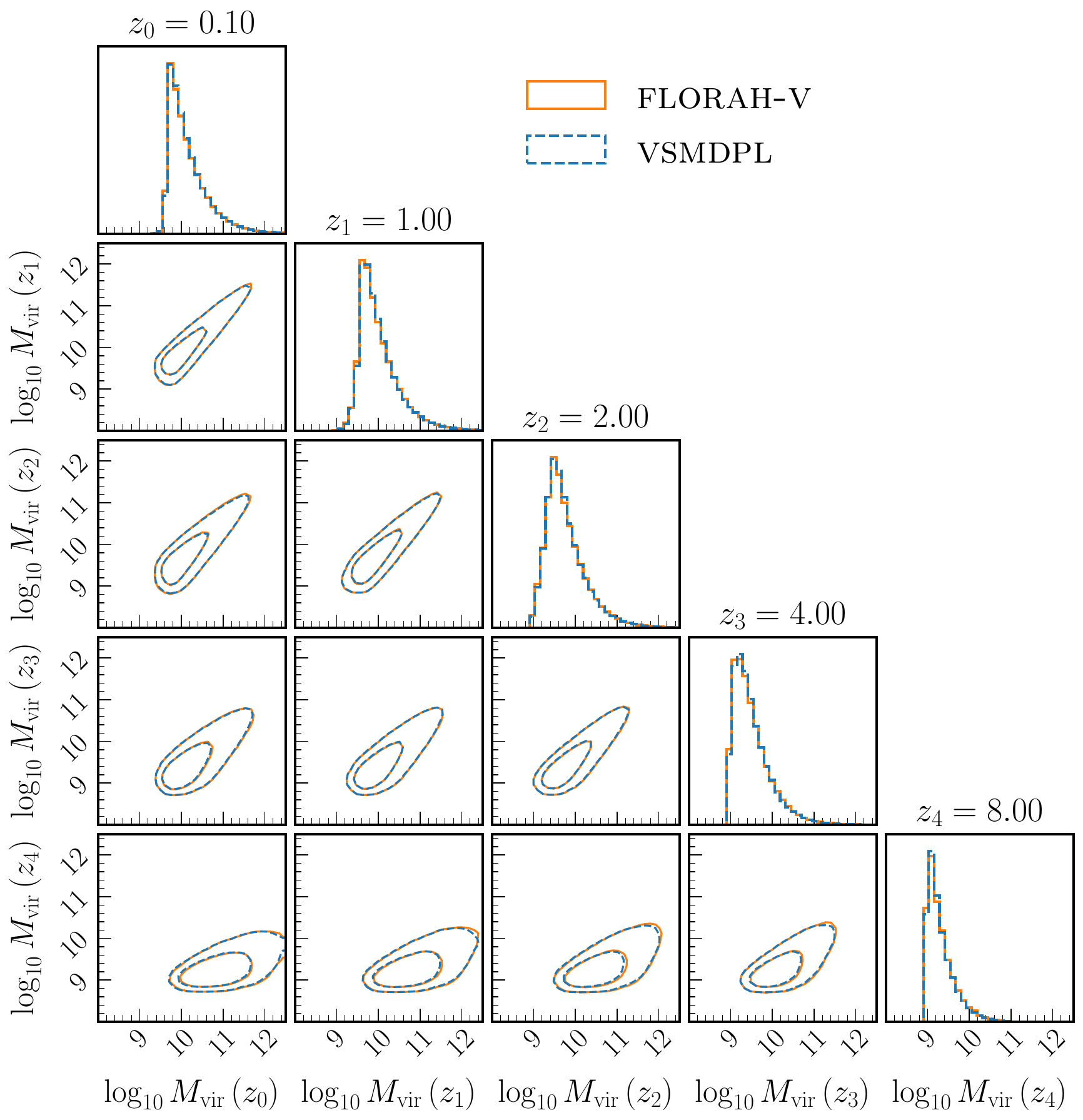}
    \includegraphics[width=0.495\linewidth]{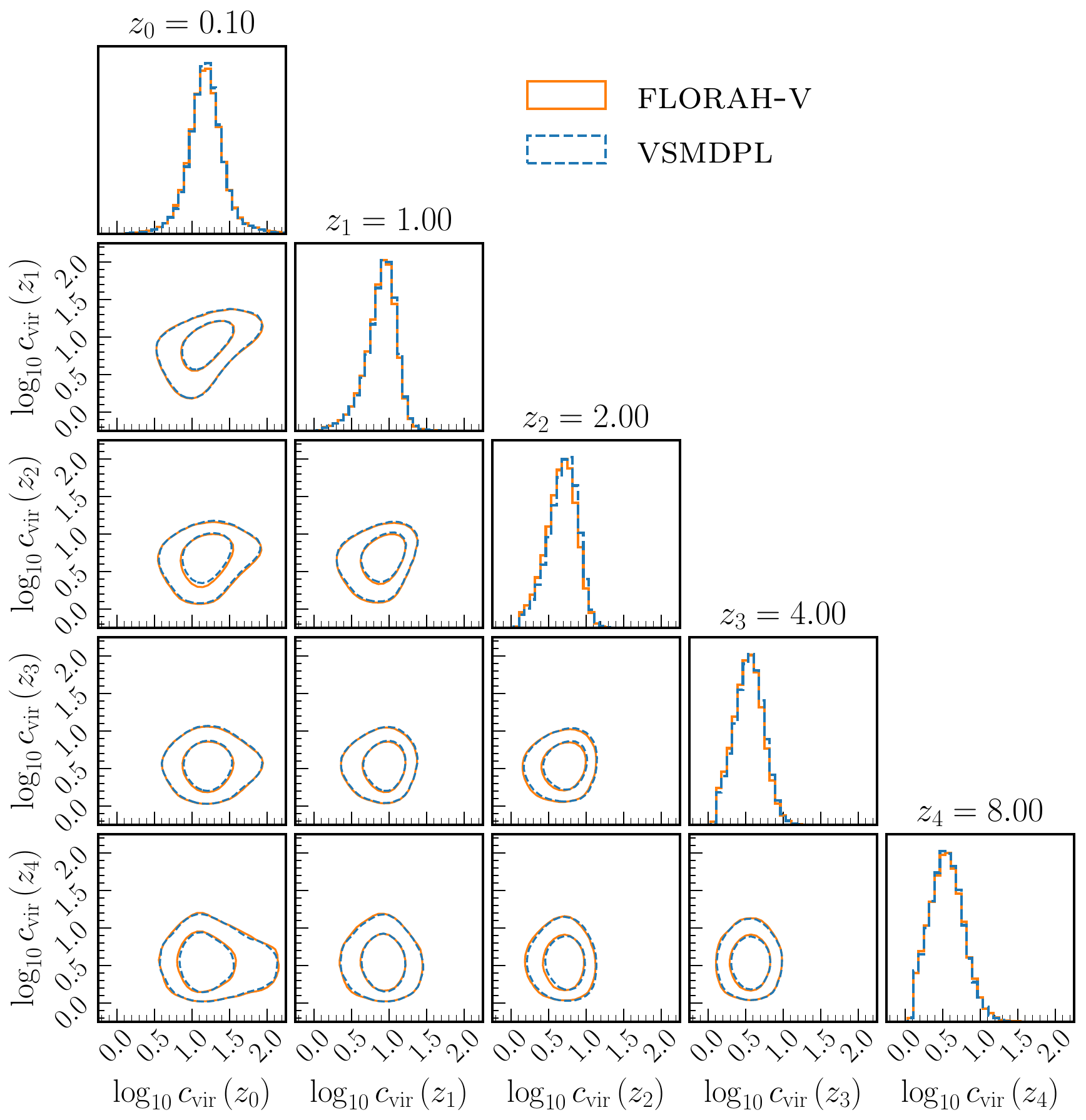}
    \caption{The joint distributions of \logMvir (left) and of \cvir (right) across a few chosen redshifts ($z=0.1, 1, 2, 4, 6, 8$) of \florahv (solid) and \vsmdpl (dashed).
    The contour lines show the 68\% and the 95\% intervals.}
    \label{fig:vsmdpl_zcorr}
\end{figure*}

\begin{figure}
    \centering
    \includegraphics[width=\linewidth]{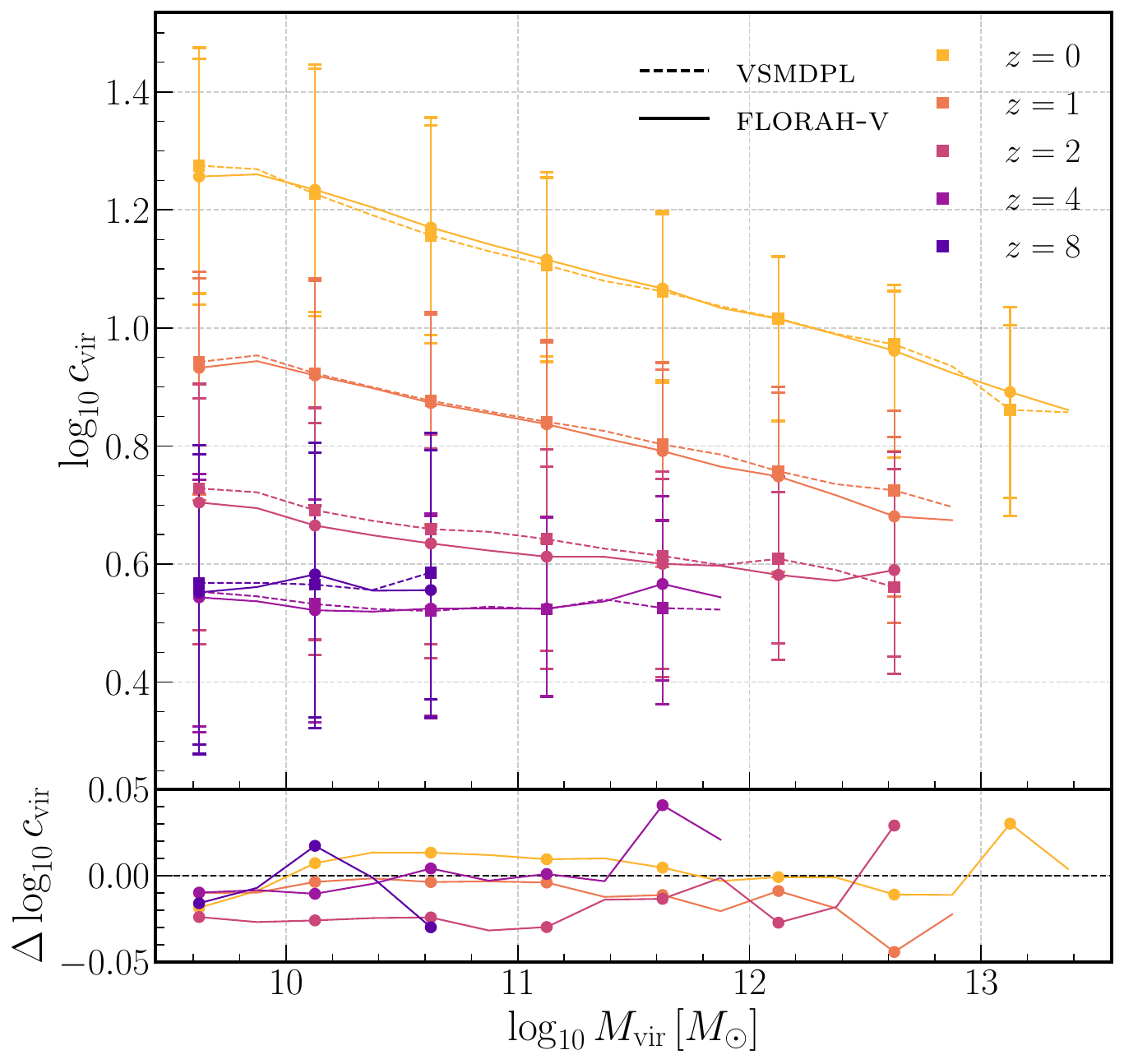}
    \caption{The median concentration-mass relations of \vsmdpl (dashed lines) and \florahv (solid lines) for five different redshift slices.
    The error bars denote the spread of the relations, computed from the 16th and 84th percentiles.
    The bottom panel shows the residual between the \florahv and \vsmdpl concentration-mass relations. 
    }
    \label{fig:vsmdpl_cvir_mass}
\end{figure}

\begin{figure}
    \centering
    \includegraphics[width=\linewidth]{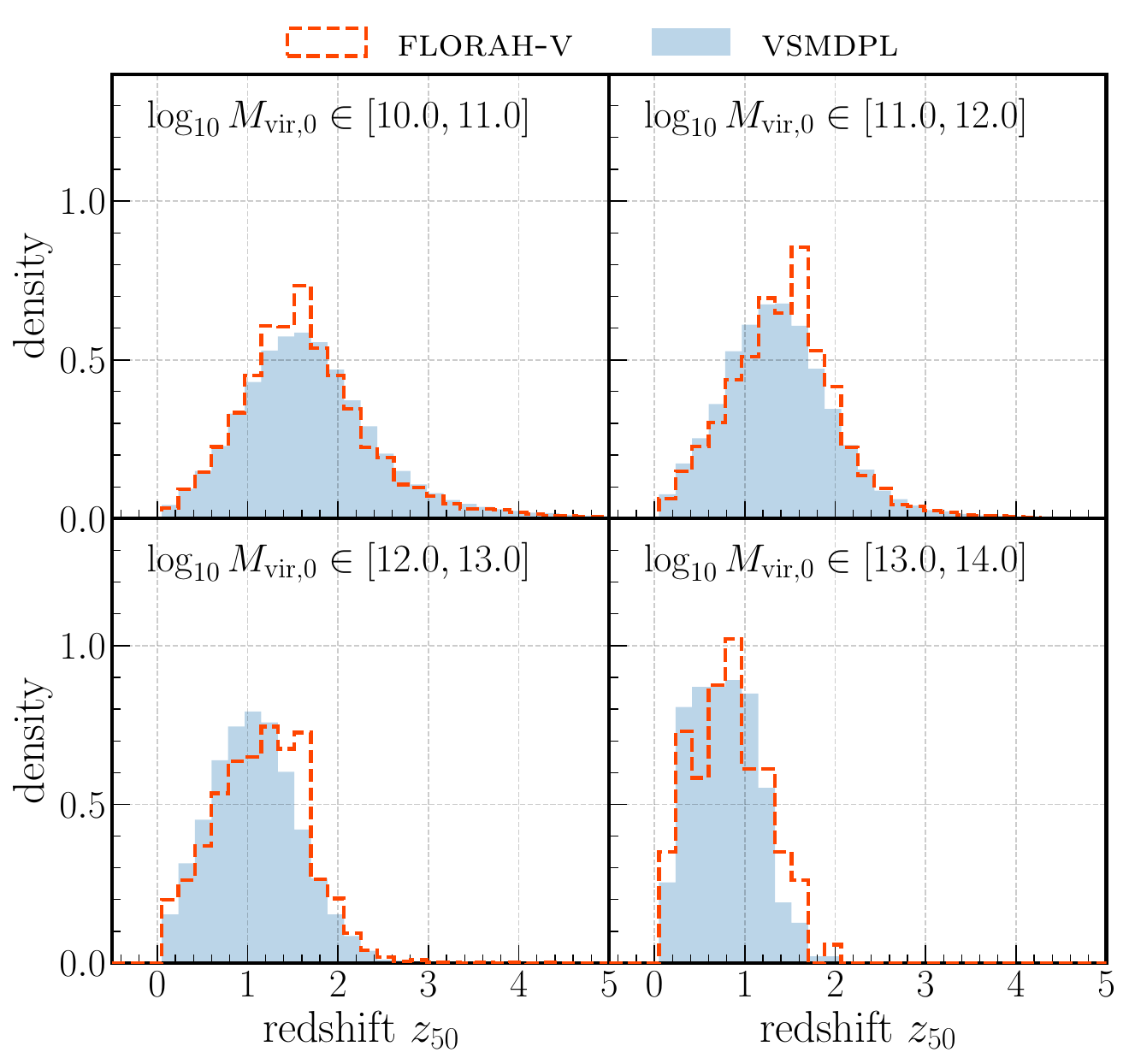}
    \caption{
    The formation redshift $z_{50}$, defined as the redshift at which the halo forms 50\% of its mass, for four mass bins of \vsmdpl.
    Masses are in \si{M_\odot} unit.
    }
    \label{fig:vsmdpl_tform}
\end{figure}

In Section~\ref{section:vsmdpl_result}, we train \florah on \vsmdpl to show results for learning MPBs of a single simulation.
Then we train \florah on the combined \gureftc simulations to show results for combining multiple simulations at different resolutions.
Finally, in Section~\ref{section:gureft_vsmdpl_result}, we combine the trained models to extend the mass and redshift coverage of \florah.

\subsection{Learning the \vsmdpl simulation}
\label{section:vsmdpl_result}

We trained \florah on \vsmdpl to test our method on its capability to learn the assembly histories of a single simulation.
We refer to this model as \florahv.
For our training and validation datasets, we extracted $306,014$ and $34,400$ MPBs, respectively, from a $(80 \, \si{Mpc \, h^{-1}})^3$ sub-volume of \vsmdpl and applied the preprocessing step in Section~\ref{section:single_box}.
The maximum training redshift is set to be $z_\mathrm{max, train}=10$.
The training process terminated at $\sim 200$ epochs or $\sim 2$ hours on a single NVIDIA Tesla V100 GPU.

For our test dataset, we extracted $387,031$ MPBs from a \emph{different} $(80 \, \si{Mpc \, h^{-1}})^3$ \vsmdpl sub-volume. 
For each MPB in the test dataset, we used the initial halo at $z=0$, taking its mass and concentration as the initial input feature $\xt{0} = (\log_{10} \Mt{0}, \ct{0})$ for the generation process. 
We sampled the scale factors every 2 -- 6 snapshots\footnote{
This choice is done out of convenience to compare the generated MPBs with the N-body simulation. In general, the time steps can be set randomly as long as they are within a reasonable range of the training time steps, as mentioned in Step (ii) of Section~\ref{section:generation}.
} starting from $z=0$ to obtain the time features of each branch ${\at{i}}$. 
We also sampled up to a redshift of $z_\mathrm{max, gen}=14$, beyond the maximum training redshift ($z_\mathrm{max, train} = 10$), to explore the model's extrapolation capability. 
As the length of each MPB is set by choosing ${\at{i}}$, which are chosen randomly, we can generate MPBs with variable lengths, and the upper bound in the number of progenitor halos depends on the minimum mass. 
We generated $387,031$ \florahv MPBs to match the number obtained from \vsmdpl. 

In Figure~\ref{fig:vsmdpl_example_all}, we show an example of a few histories generated by \florahv.
Each column shows the mass (top panel) and concentration (bottom panel) histories of an example root halo in the \vsmdpl simulation (blue).
For each root halo, we use \florahv to generate 30 different realizations of the histories (orange).
The shaded regions denote the extrapolation region beyond $z_\mathrm{max, train} = 10$, as mentioned above.
This figure demonstrates the stochasticity of the assembly histories of halos: for the same descendant mass and concentration, we can arrive at vastly assembly scenarios. 
In addition, we see that \florah-generated MAHs can capture both smooth accretion, characterized by a steady increase in the MAHs, as well as implicit merger events, characterized by sharp jumps in the MAHs. 
This highlights the advantage of \florah over other approaches, distinguishing it from other methods that predominantly capture the average trends of MAH.

\begin{figure*}
    \centering
    \includegraphics[width=\linewidth]
    {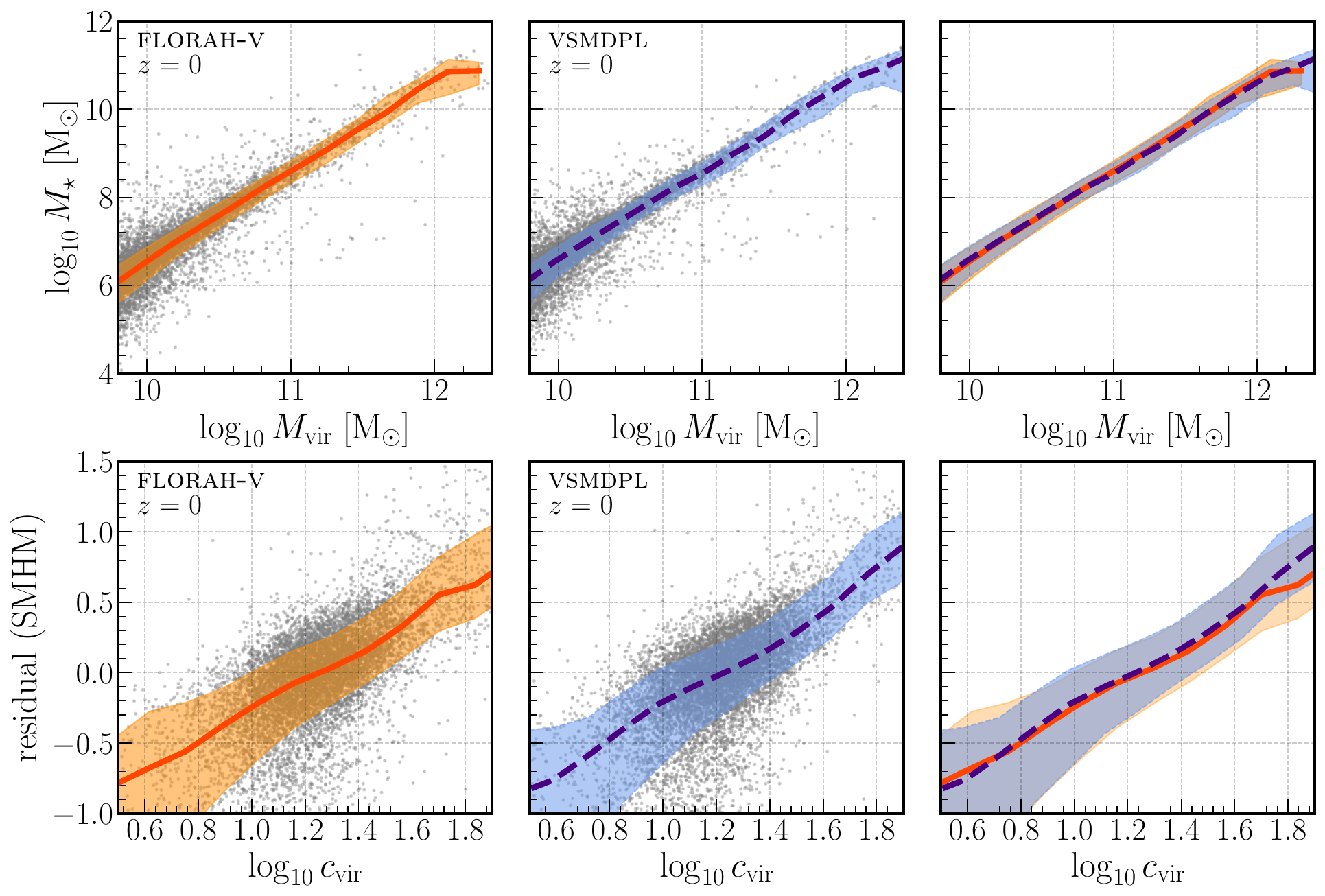}
    \caption{    
    Top: The stellar-to-halo mass relation (SHMR) at $z=0$ computed by the SC-SAM. 
    Bottom: The residual of SHMR, defined as the difference between the $M_\star / M_\mathrm{vir}$ value for each halo and the median value in its corresponding \Mvir bin, as a function of the DM halo concentration. 
    In both panels, the shaded regions represent the middle-68 \% containment regions. 
    }
    \label{fig:vsmdpl_sam}
\end{figure*}

\begin{figure*}
    \centering
    \includegraphics[width=0.9\linewidth]{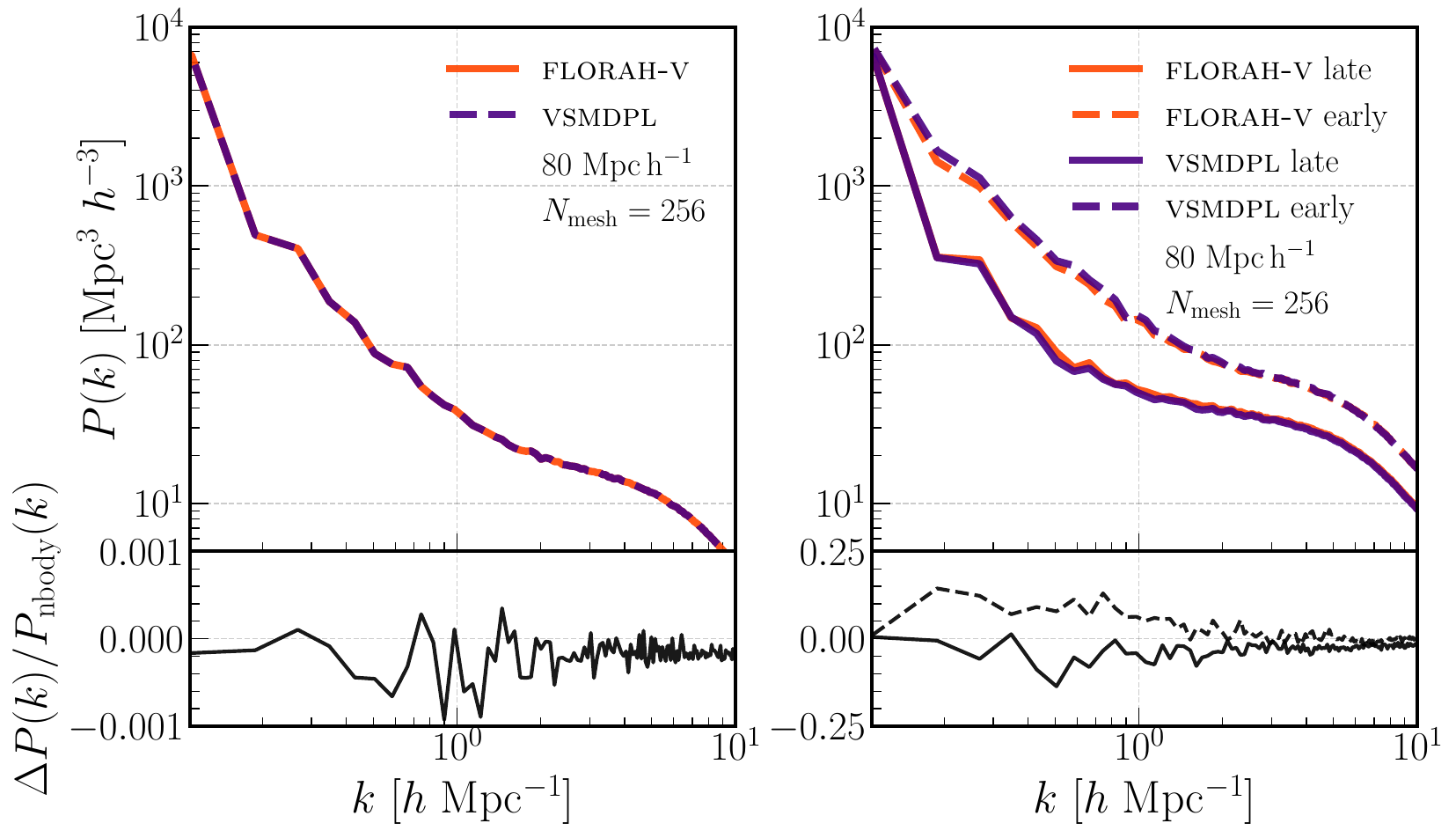}
    \caption{
    Left: The top panel shows the galaxy power spectra $P(k)$ at $z=0$ of galaxy catalogs from \vsmdpl (dashed blue) and generated by \florahv (solid orange).
    The bottom panel shows the ratio of the residual $\Delta P(k) = P_\florah(k) - P_\vsmdpl(k)$ and the \vsmdpl power spectrum. 
    Right: The top panel shows the galaxy power spectra $P(k)$ at $z=0$ of galaxy catalogs from \vsmdpl (blue) and generated by \florahv (orange) divided into a late-forming galaxy population (solid) and an early-forming population (dashed).
    The bottom panel shows the ratio of the residual $\Delta P(k)$ and the \vsmdpl power spectrum for the corresponding population.
    Similarly, the solid line represents the late-forming population, while the dashed line represents the early-forming population.
    \label{fig:vsmdpl_ps}
    }
\end{figure*}

\subsubsection{Progenitor-descendant ratio distribution}
\label{section:vsmdpl_result_dlogm}
In the following sections, we present various tests to validate the mass and concentration histories generated by \florahv.
In Figure~\ref{fig:vsmdpl_dlogm}, we show the distribution of the progenitor-descendant mass ratios $\log_{10} M_P/M_D$ across four mass bins (columns) and three redshift slices (rows).
The distribution of $\log_{10} M_P/M_D$ is learned directly by \florahv during the training process and thus presents a good first validation test.
We see good agreement between the distributions from \vsmdpl (shaded blue) and \florahv (solid orange).
Note that because the last mass bin with $\log_{10}{M_\mathrm{vir, 0}} \in \left[13.0, 14.0\right] \, \si{dex}$ contains only 373 MPBs, the distributions are noticeably noisier.
In each bin of the first row of Figure~\ref{fig:vsmdpl_dlogm}, the descendant halos are also the root halos and thus have roughly the same virial mass. 
However, we see that the distribution of the progenitor-descendant mass ratios can span a wide range of values, which further demonstrates the stochasticity of the MAHs between redshifts.

\subsubsection{Mass and DM concentration histories}
\label{section:vsmdpl_result_mah}

In Figure~\ref{fig:vsmdpl_mah_cah}, we show the median, middle-68\% percentile, and middle-95\% percentile containment region of the assembly histories of \vsmdpl (blue) and \florah (orange), along with their residuals, in four mass bins.
We normalize the MAHs by dividing the progenitor masses by the root mass so that they can be more easily compared between bins. 
The residual of the containment region is computed by averaging the residuals of the corresponding upper and lower percentile curves.
As can be seen from the slopes of the MAHs, low-mass halos tend to form earlier than high-mass halos, as has been observed in previous studies of galaxy formation (e.g.~\citep{2001astro.ph.11069W}).
The MAHs plateau out at around the resolution limit $M_\mathrm{min, halo} = 100 \, M_\mathrm{DM}$ (dashed black line, computed from the high end of each mass bin), as expected, because we remove all halos below this mass limit.
In all mass bins, the MAHs predicted by \florahv are consistent with those from \vsmdpl.
As mentioned above, the last mass bin $\log_{10}{M_\mathrm{vir, 0}} \in \left[13.0, 14.0\right] \, \si{dex}$ contains significantly fewer MPBs, so the containment regions are noisier.
In the bottom panels of Figure~\ref{fig:vsmdpl_mah_cah}, we show the DM concentration \cvir histories of \vsmdpl and \florah.
Similarly, the concentration histories predicted by \florah agree well with those of \vsmdpl.
The shaded gray box represents the $z > z_\mathrm{max, train}=10$ region, beyond which the MAHs and concentration histories are extrapolated. 
Except for the first bin, \florahv demonstrates an impressive capability to extrapolate and predict the MAHs and concentration histories up to $z=14$. 
In the lowest mass bin, \florah accurately extrapolates up to about $z \approx 12$, but it overestimates the MAHs after this point. 
We attribute this to the fact that this bin is the closest to the resolution limit $M_\mathrm{min, halo}$, as can be seen from the horizontal dashed line, and caution against extrapolating \florah near this limit.

In Figure~\ref{fig:vsmdpl_zcorr}, we show the joint distribution between a few chosen redshifts ($z=0.1, 1, 2, 4, 6, 8$) for \Mvir and \cvir for \florahv-generated (solid orange) and \gureftc (dashed blue) MPBs.
The 68\% and 95\% of the \Mvir joint distributions are shown in the left panel and \cvir joint distributions in the right panel.
The \Mvir values at different redshifts are highly and positively correlated.
The mass correlation becomes weaker over a wider redshift range, as expected. 
On the other hand, the \cvir values are more correlated at low redshifts (although still much less than \Mvir) and uncorrelated at high redshifts. 
For both \Mvir and \cvir, \florahv can learn the joint distributions and correlations between redshifts present in \vsmdpl simulation.

We also emphasize that \florah jointly predicts the mass and DM concentration of the halos at each redshift using the conditional normalizing flows.
The flows excel at modeling high-dimensional correlations, so additional halo properties (e.g., halo shape, environment density) can naturally be incorporated into \florah \textit{without requiring additional assumptions about the data. 
In Figure~\ref{fig:vsmdpl_cvir_mass}, we show the median concentration-mass relations at different redshifts, as derived from \vsmdpl (dashed lines) and \florahv-generated (solid lines) MPBs. 
The error bars represent the spread in the concentration-mass relations.
The bottom panel shows the residual between the \florahv and \vsmdpl median. 
In general, we see that the concentration-mass relations recovered by \florahv are consistent with \vsmdpl across multiple redshifts. }

In Figure~\ref{fig:vsmdpl_tform}, we show the distribution of the formation redshift $z_{50}$, defined as the redshift at which the halo forms 50\% of its mass, for the four chosen mass bins. 
As mentioned above, low-mass halos tend to form at higher redshift than high-mass halos. 
We see that the $z_{50}$ distribution by \florahv agrees well with \vsmdpl.
We emphasize that the formation redshift $z_{50}$ is not explicitly learned by the network but instead derived from the MAHs. 
This further shows that \florah can capture the ``long-range'' mass correlation in the MAHs.

\subsubsection{Observables and Halo assembly bias}
\label{section:assembly_bias}

We do not directly observe dark matter halos or merger trees, but only the observable properties of galaxies.
In semi-analytic models (SAMs), these properties are predicted by using merger trees as inputs and solving ordinary differential equations for observables like galaxy luminosities and quasi-observables like stellar mass.
Thus we need to test that \florah-generated assembly histories can be used in place of N-body assembly histories in SAMs to predict these observables. 
To that end, here we apply the SC-SAM to predict the stellar masses $M_\star$ of galaxies. 
For a fair comparison with \vsmdpl, we input only the MPBs of \vsmdpl trees to the SAM.
We use both the mass and DM concentration of the generated MPBs.
In SC-SAM, the concentration is used in two ways: first, in computing the galaxy sizes (see~\citealt{somerville2008a}), and second, in computing the timescale for disruption of satellite galaxies by tidal forces.
Note that since we do not model satellite galaxies in our analysis, the latter point is not directly relevant.
We apply the minimum progenitor mass cut $M_\mathrm{min, halo}$ to both \vsmdpl and \florah halos. 
In addition, we do not restrict the redshift range of either \vsmdpl and \florah halos.  

In the top panel of Figure~\ref{fig:vsmdpl_sam}, we show the stellar-to-halo mass relation (SHMR).
Next, we compute the SHMR residual, defined as the difference between the $M_\star/M_\mathrm{vir}$ value for each halo and its median value in the corresponding \Mvir bin.
The correlation between the SHMR residual and the DM concentration \cvir is shown in the bottom panel.
Estimating the residual this way removes most of the mass dependence. 
This demonstrates that the galaxy properties depend on secondary halo characteristics beyond halo mass, which can lead to the phenomenon known as ``assembly bias'' \citep{Hadzhiyska2021, 10.1093/mnras/stac2297}. 
In both cases, relations predicted by \florah are consistent with \vsmdpl.
\emph{A SAM run on \florah-generated assembly histories accurately reproduces the correlation between stellar mass, halo formation history, and halo concentration.} 

However, in the top panel of Figure~\ref{fig:vsmdpl_sam}, we observe a bifurcation in the SHMR, leading to a secondary population of galaxies with low stellar mass.
We theorize that this population represents galaxies which normally grow their mass via mergers. 
Since we only input MPBs into the SC-SAM, we underestimate the stellar mass of these galaxies.
The bifurcation disappears when we input full merger trees of \vsmdpl into the SC-SAM, which further supports our theory.
Currently \florah only generates MPBs, and thus it is unable to capture these merger-driven galaxies.
We have also found that using only the MPBs, SC-SAM does not correctly recover the mass of the central supermassive black holes for both \vsmdpl and \florah. 
This is expected because the central supermassive black holes grow mass via mergers.
We further discuss the limitations of the current framework and future outlook for \florah in Section~\ref{section:discussion}.

\subsubsection{Clustering}
\label{section:vsmdpl_clustering}

To further probe the assembly bias, we compute the galaxy power spectra $P(k)$ of \vsmdpl and \florahv-generated galaxy catalogs at $z=0$.
\florah does not generate positions, and so for each halo, we take the positions of the corresponding \vsmdpl halo in the test dataset at $z=0$.
Thus, only the stellar masses computed by SC-SAM differ between the two catalogs.
In the top left panel of Figure~\ref{fig:vsmdpl_ps}, we show the galaxy power spectra $P(k)$ of \vsmdpl (dashed blue) and \florahv (solid orange) catalogs.
The ratio between the residual
$\Delta P(k) = P_\florah(k) - P_\vsmdpl(k)$ and the \vsmdpl power spectrum is shown in the bottom panel.
We see excellent agreement between the power spectrum of the \florahv catalog and that of the \vsmdpl catalog. 
We emphasize that this test is non-trivial because \florah is trained on individual halos, independently, and has no notion of any spatial correlations when generating MAHs for a population of clustered halos together.  

Studies of assembly bias have also shown that early-forming galaxies exhibit stronger clustering tendencies compared to their late-forming counterparts (e.g., \cite{2006ApJ...652...71W}).
To investigate this phenomenon, we categorized each galaxy catalog into two groups based on their formation redshift $z_{50}$: a late-forming population comprised of the first 25\% of halos, and an early-forming population consisting of the last 25\%. 
The resulting power spectra $P(k)$ and residuals are displayed in the right panels of Figure~\ref{fig:vsmdpl_ps}, with the solid line representing the late-forming population and the dashed line representing the early-forming population.
The orange lines represent \florahv galaxies, while the blue lines represent \vsmdpl galaxies.
Our analysis indicates that the power spectra of \florahv galaxies are consistent with those of \vsmdpl galaxies for both populations. 
This further suggests that \florah-generated histories can accurately replicate the assembly bias of galaxies.

\subsection{Learning the combined \gureft simulations}
\label{section:gureft_result}

Using the procedure in Section~\ref{section:combine_box} to combine the four \gureft boxes, we extracted $42,774$ unique MPBs and applied the data augmentation steps in Section~\ref{section:single_box}.
Note that \gureftc contains significantly fewer halos than \vsmdpl because the \gureft boxes have much smaller volumes and particle numbers (as can be seen from Table~\ref{tab:sim}).
For this reason, we used $38,023$ MPBs for the training dataset and $4,751$ MPBs for the validation dataset and did not create a test dataset for \gureftc.
We trained a model, denoted as \florahg, up to a redshift of $z_\mathrm{max, train}=20$.
Then we generated $4,751$ \florahg MPBs to match the number obtained from \gureftc. 
During the generation process, we applied a similar procedure in Section~\ref{section:vsmdpl_result} to choose the starting halo $\xt{0} = (\log_{10} \Mt{0}, \ct{0})$ at $z=5.89$ and list of scale factors from the \textit{validation} dataset. 
Similarly, we generate MPBs up to a redshift $z_\mathrm{max, gen}=24 > z_\mathrm{max, train}$ to demonstrate the model's extrapolation capability.

\begin{figure*}
    \centering
    \includegraphics[width=0.9\textwidth]{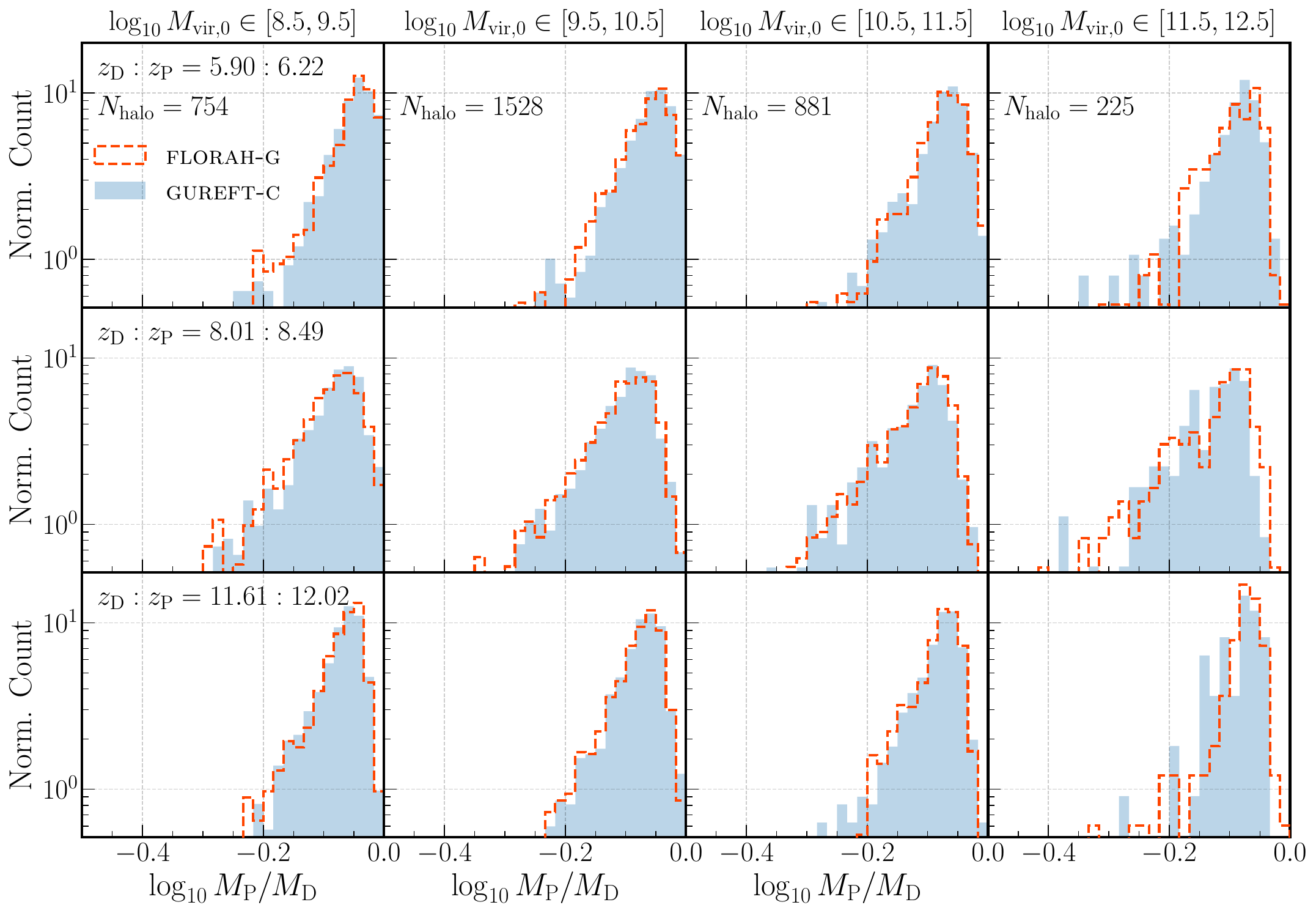}
    \caption{
    The normalized distribution of the progenitor-descendant mass ratio $\log_{10} M_P/M_D$ as modeled by \florahg (blue dashed line) and from the \gureftc simulation (shaded orange).
    Panels are the same with Figure~\ref{fig:vsmdpl_dlogm}.
    }
    \label{fig:gureftc_dlogm}
\end{figure*}

\begin{figure*}    
    \centering
    \includegraphics[width=\linewidth]{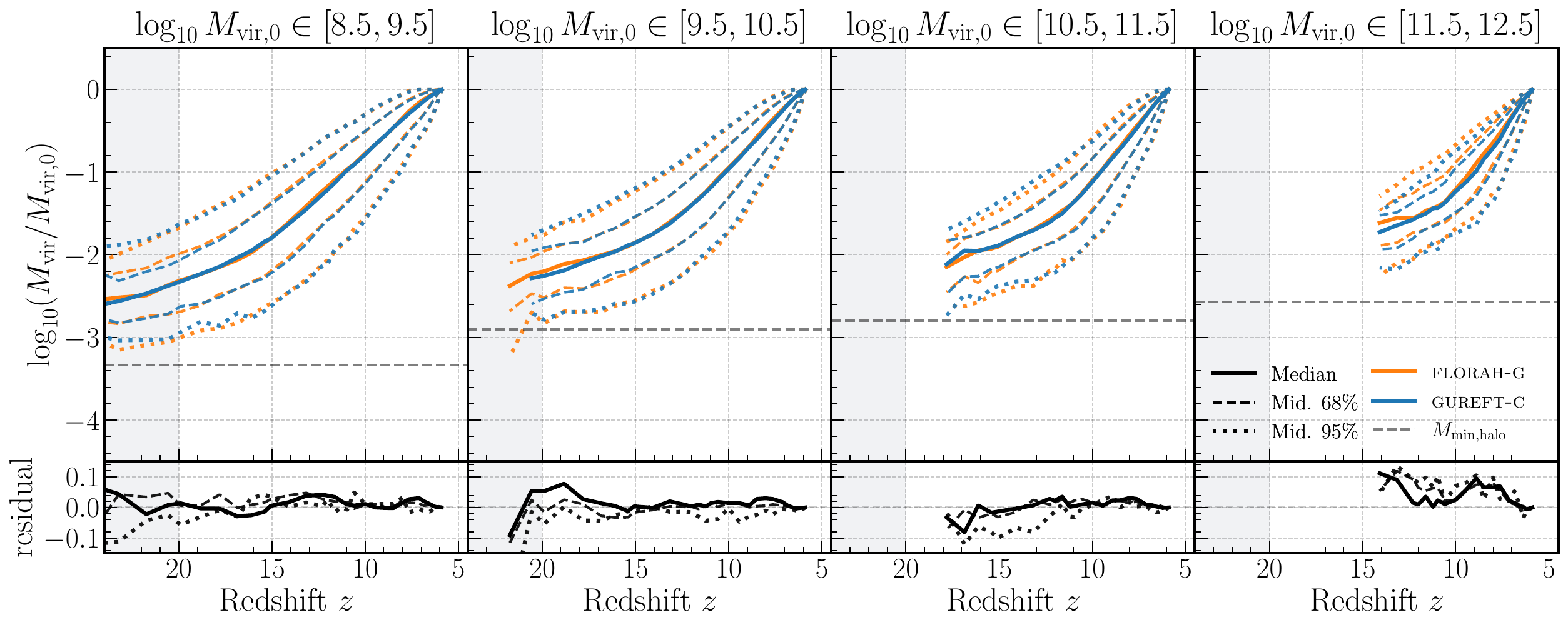}
    \includegraphics[width=\linewidth]{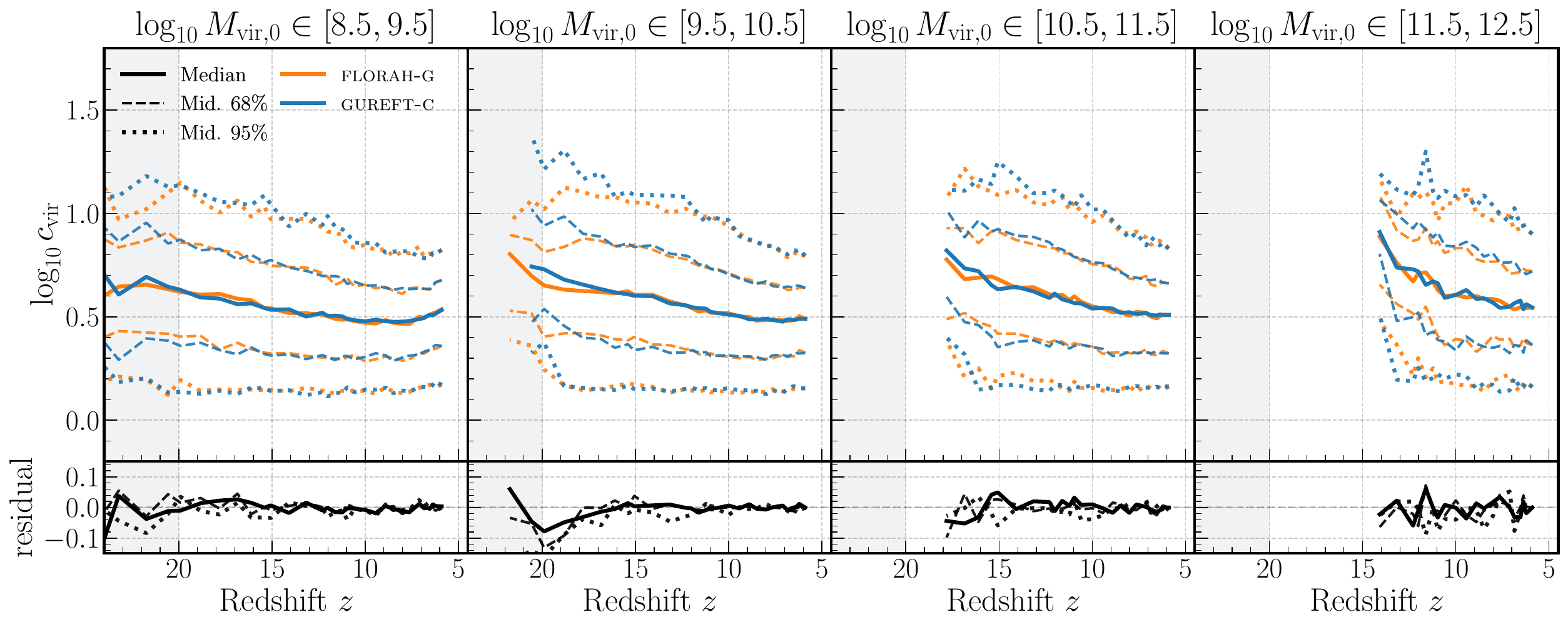}
    \caption{
    The MAHs and DM concentration histories of the MPBs in \gureftc and \florahg in four mass bins (in \si{M_\odot unit}). Panels are the same as Figure~\ref{fig:vsmdpl_mah_cah}.
    }
    \label{fig:gureft_mah_cah}
\end{figure*}

\begin{figure*}
    \centering
    \includegraphics[width=0.495\linewidth]{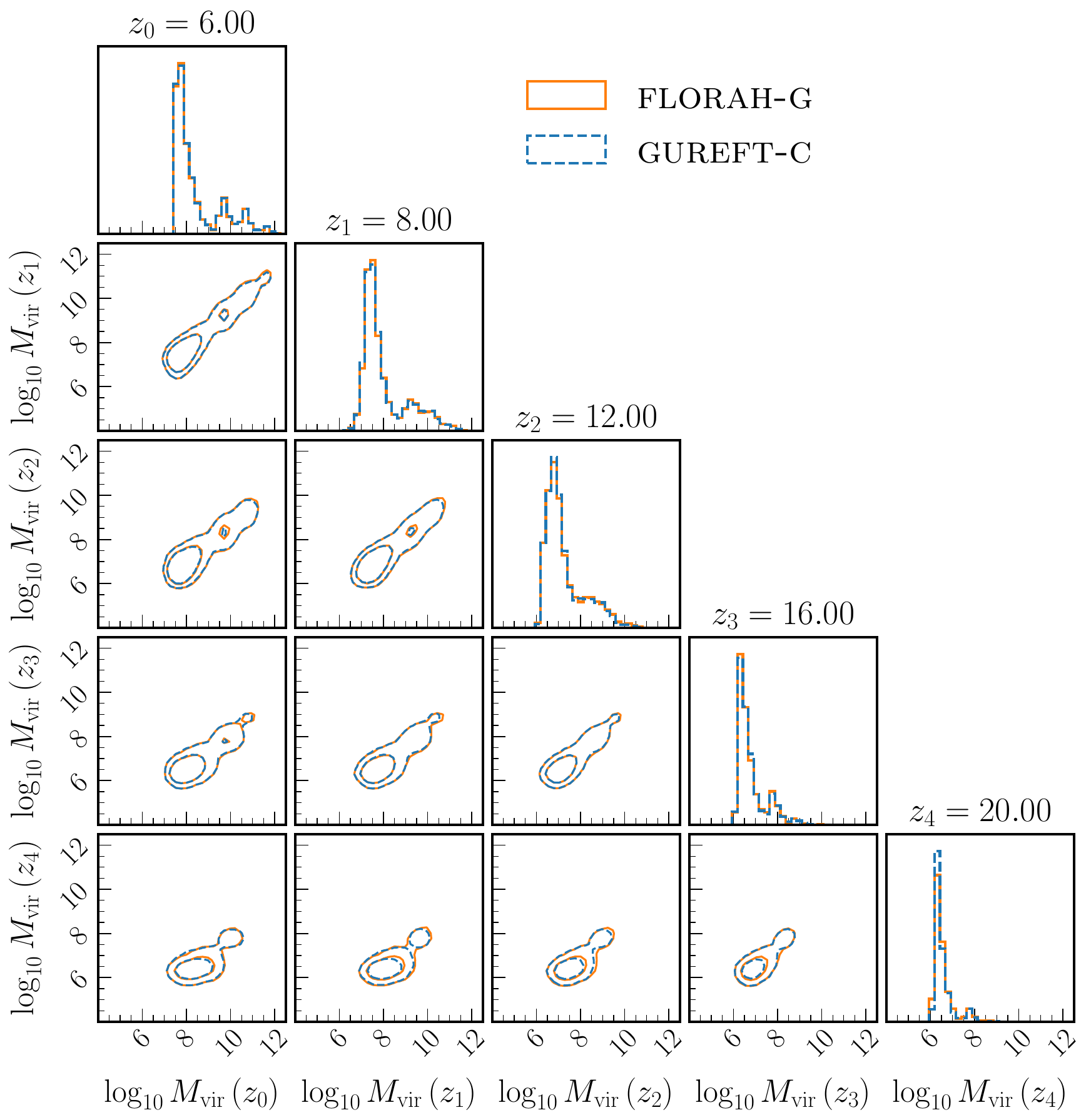}
    \includegraphics[width=0.495\linewidth]{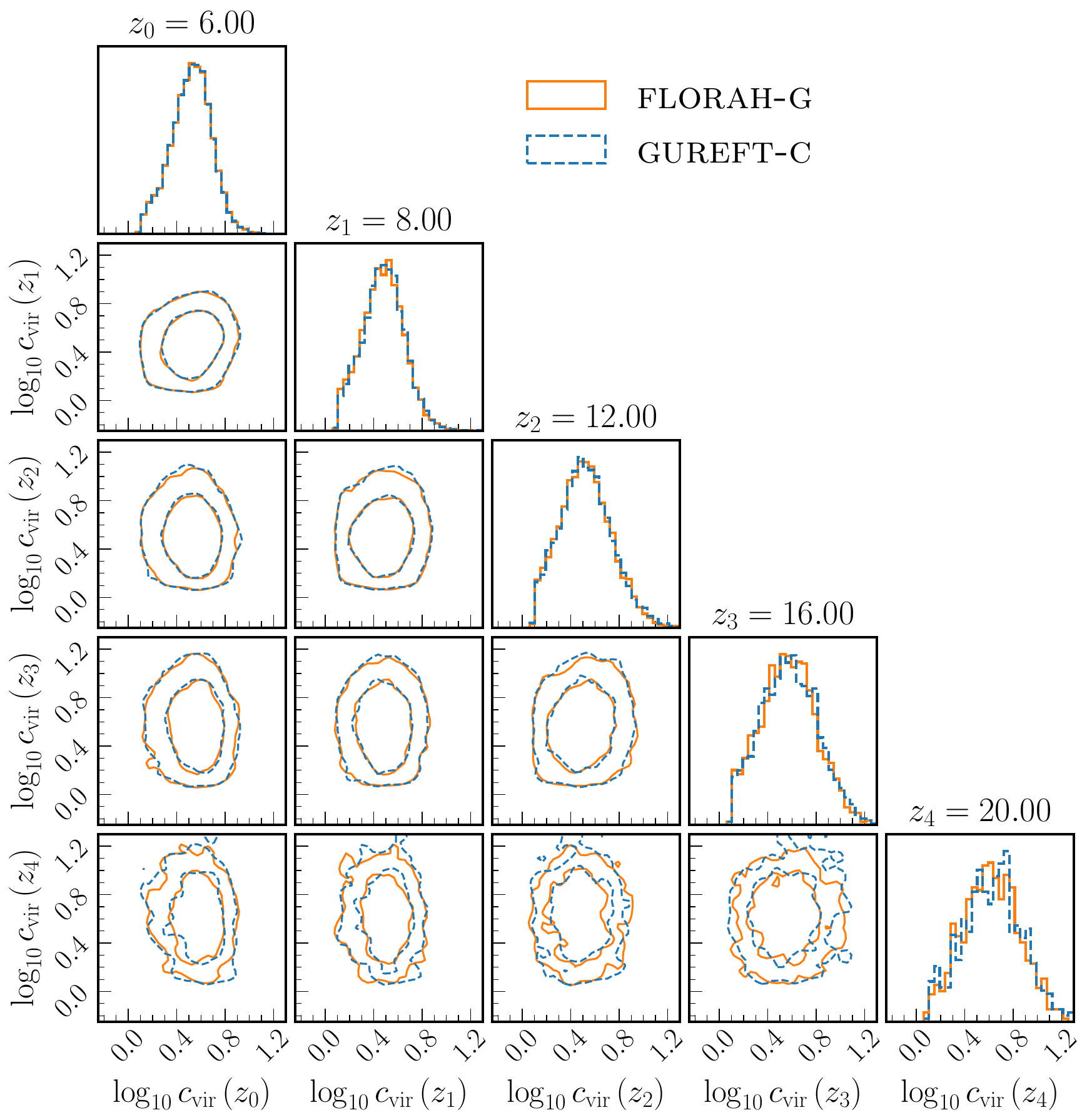}
    \caption{The joint distributions of \logMvir (left) and \cvir (right) across a few chosen redshifts ($z=6, 8, 12, 16, 20$) of \florahg (solid) and \gureftc (dashed).
    The contour lines show the 68\% and the 95\% intervals.}
    \label{fig:gureft_zcorr}
\end{figure*}

We perform additional validation checks similar to Section~\ref{section:vsmdpl_result}.
In Figure~\ref{fig:gureftc_dlogm}, we compare distributions of the progenitor-descendant mass ratio $\log_{10} M_D / M_P$ from \gureftc (shaded blue) and generated by \florahg (solid orange) for four mass bins (columns) and three redshift slices (rows).
We choose the four mass bins such that each bin corresponds to a different \gureft box, with \gureft-05 being the least massive bin and \gureft-90 being the most massive bin. 
Similar to the \vsmdpl example in Section~\ref{section:vsmdpl_result}, we observe good agreement between the distributions from \gureftc and generated by \florahg.

In Figure~\ref{fig:gureft_mah_cah}, we present the median, 68\%-percentile, and 95\%-percentile containment regions of the MAHs (top panels) and the DM concentration histories (bottom panels) for MPBs from \gureftc (blue) and generated by \florahg (orange). 
Similarly to \vsmdpl, low-mass halos tend to form at earlier redshifts than high-mass halos.
Also as expected, the MAHs of both \gureftc and \florahvg plateau out at $M_\mathrm{min, halo}$.
There are a few differences compared to \vsmdpl and the low-redshift case.
Unlike in \vsmdpl, where $M_\mathrm{min, halo}$ is fixed at 100 times the DM mass of \vsmdpl, here $M_\mathrm{min, halo}$ is determined by the $\log_{10} M_\mathrm{vir, 0}$ at $z=5.89$ (refer to Table~\ref{tab:sim}).
We choose the same four mass bins as Figure~\ref{fig:gureftc_dlogm}, with each bin corresponding to a different \gureft box.
Due to a low number of halos in the validation dataset, we only display redshift slices with more than 10 resolved halos, resulting in the termination of the MPBs before $z_\mathrm{max, train}=20$ in the last two mass bins. 
Again, we observe good agreement between the MAHs and concentration histories of \florahg MPBs with those of \gureftc MPBs.
Moreover, \florahg exhibits a similar capability to extrapolate the MPBs beyond $z_\mathrm{max, train}=20$ in the first two mass bins.

In Figure~\ref{fig:gureft_zcorr}, we show the joint distributions of \Mvir (left panels) and \cvir (right panels) at selected redshifts ($z=6, 8, 12, 16, 20$) for \florahg-generated (solid orange) and \gureftc (dashed blue) MPBs. 
The contour lines represent the 68\% and 95\% intervals. 
Once again, the distributions of \florahg MPBs match well with those of \gureftc MPBs. 
Note that the mass distribution of \gureftc exhibits multiple peaks due to the applied mass cuts in the procedure to combine the four \gureftc boxes (see Section~\ref{section:combine_box}).

\begin{figure}
    \centering
    \includegraphics[width=\linewidth]{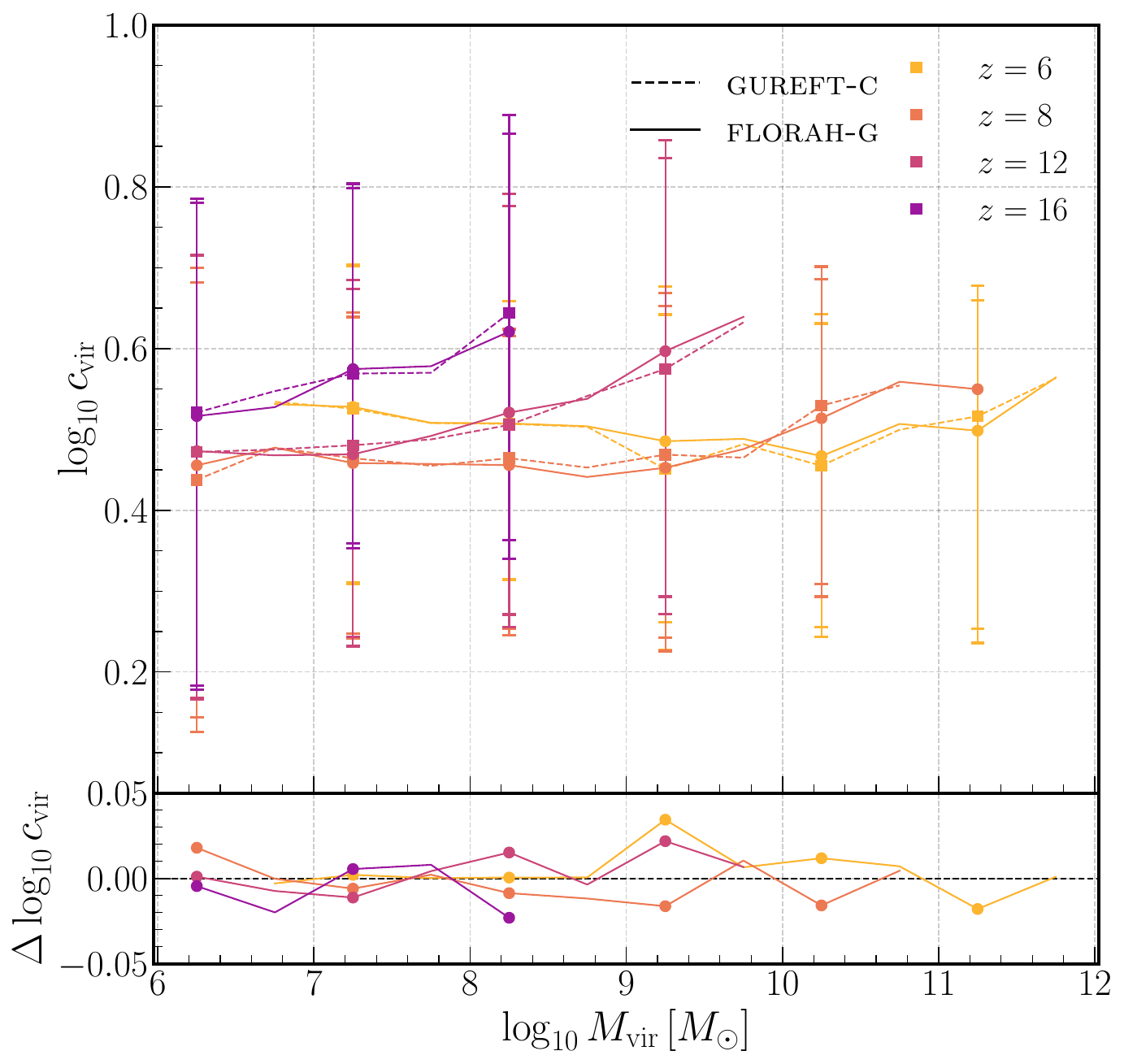}
    \caption{The median concentration-mass relations derived from \gureftc (dashed lines) and \florahg-generated MPBs (solid lines). Panels are the same with Figure~\ref{fig:vsmdpl_cvir_mass}}
    \label{fig:gureft_cvir_mass}
\end{figure}

In Figure~\ref{fig:gureft_cvir_mass}, we show the median concentration-mass relations from the combined \gureftc boxes and \florahg-generated MPBs.
Similar to the \vsmdpl example in Figure~\ref{fig:vsmdpl_cvir_mass}, the error bars denote the spread of the relations, computed from the 16th and 84th percentiles.
The bottom panel shows the residual between \florahg and \gureftc median relations. 
Here, we also see that the relations between \gureftc and \florahg agree well.

\begin{figure}
    \centering
    \includegraphics[width=\linewidth]{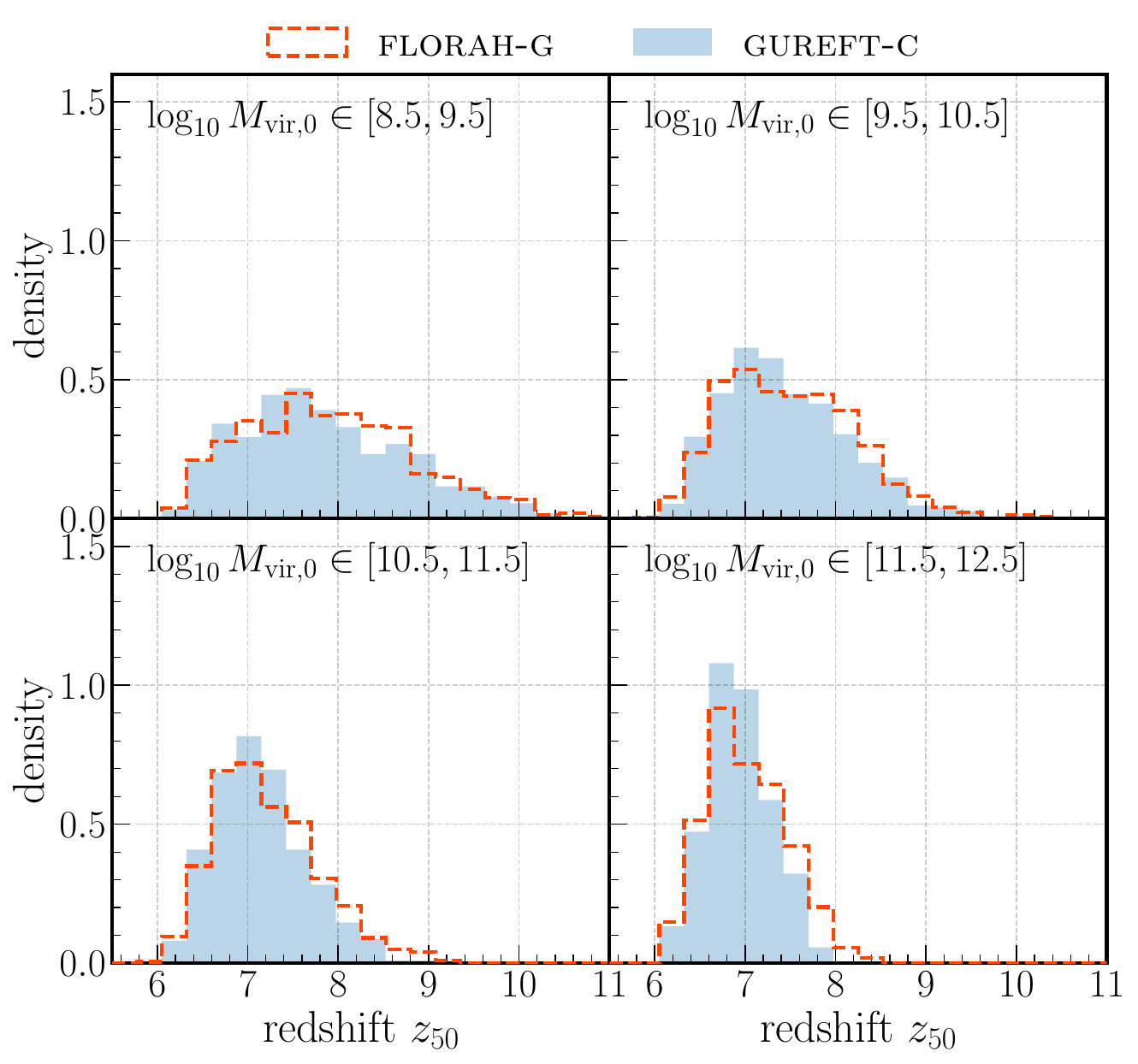}
    \caption{
    The formation redshift $z_{50}$,  defined as the redshift at which the halo forms 50\% of its mass, for four mass bins of \gureftc.
    Masses are in \si{M_\odot} unit.
    }
    \label{fig:gureft_tform}
\end{figure}

Finally, we show the distribution of the formation redshift $z_{50}$ for the four chosen mass bins of \gureftc in Figure~\ref{fig:gureft_tform}.
We see that similarly as in the \vsmdpl case, we recover the $z_{50}$ distribution of the \gureftc simulations.
Again, we emphasize that the formation redshift $z_{50}$ is a derived quantity from the MAHs and not learned directly by the network, indicating that \florah can capture the ``long-range'' mass correlation in the MAHs.
We opt not to apply the same SAM analysis shown in Section~\ref{section:assembly_bias} and \ref{section:vsmdpl_clustering} on \florahg and \gureftc MPBs.
Given that \florahg can recover the environment-dependent quantities such as the $z_{50}$, we expect the SHMR, SHMR residual-concentration relation, and galaxy clustering predicted by \florahg to also be consistent with the simulations.

To conclude this section, we developed a procedure to combine multiple N-body simulations with varying resolutions into the training dataset and showed that \florah can accurately learn the MAHs and concentration histories of these simulations. 
The \gureft simulations in this example are designed to capture the assembly histories of DM halos at the ultra-high redshifts at an unprecedented temporal resolution.
By successfully incorporating the \gureft simulations into our training dataset, \florah becomes an essential tool for understanding and analyzing the formation and evolution of cosmic structures during the early Universe. 

\subsection{Combining \gureft and \vsmdpl}
\label{section:gureft_vsmdpl_result}

\begin{figure*}
    \centering
    \includegraphics[width=0.9\linewidth]{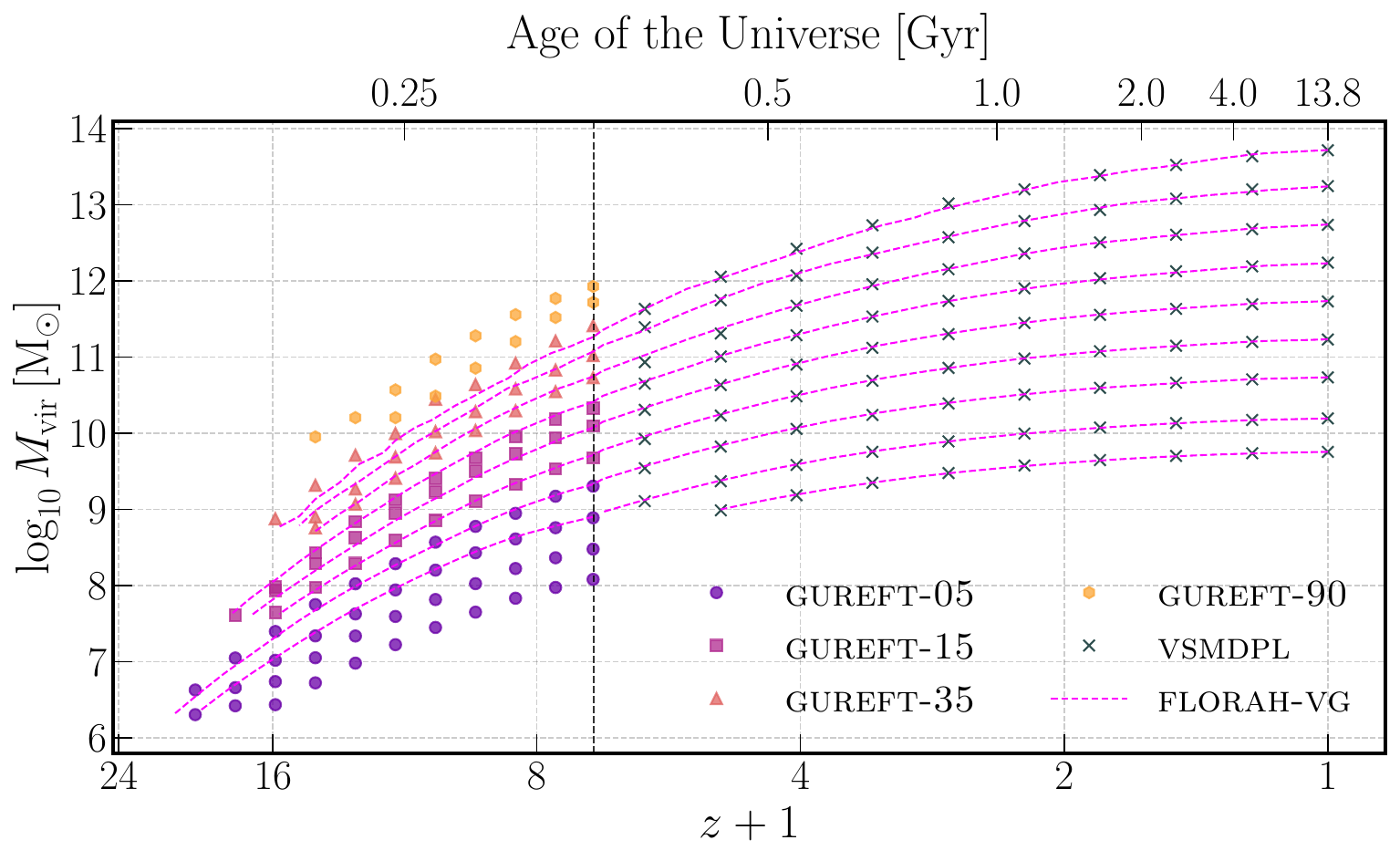}
    \caption{
    The median MAHs of the MPBs of \gureft, \vsmdpl, and \florahvg down to $M_\mathrm{min, halo} = 100 \, M_\mathrm{DM}$, where $M_\mathrm{DM}$ is the mass of the DM particle in the corresponding simulations.
    The dotted vertical line denotes the transition point from \vsmdpl to \gureftc at $z=5.89$.}
    \label{fig:mean_mah_sim_florah}
\end{figure*}

High computational costs make it infeasible to run N-body simulations at the mass resolution required to simultaneously capture the merger dynamics from dwarf galaxies ($10^5 - 10^{10}~\si{M_\odot}$) to galaxy clusters ($10^{14} - 10^{15}~\si{M_\odot}$) up to high redshifts and in large volumes.
The \gureft simulations are simulated up to $z \approx 6$ and designed to be complementary to past large volume simulations such as \vsmdpl.
In this section, we combine the previously trained model, \florahv on \vsmdpl and \florahg on \gureftc, in an attempt to generate MPBs spanning $z = 0$ to $z \approx 24$.

The generation of the combined assembly histories is done in two steps.
First, we applied a similar procedure in Section~\ref{section:vsmdpl_result} to choose the starting halo $\xt{0} = (\log_{10} \Mt{0}, \ct{0})$ at $z=0$ from the test dataset of \vsmdpl.
As before, we chose the list of \vsmdpl scale factors by sampling every 2 -- 6 \vsmdpl snapshots from $z=0$ and used \florahv to generate the MPBs.
Here, we only generate the MPBs up to a redshift of $z=5.89$ and down to a progenitor mass of $M_\mathrm{min, halo} = 5.89 \times 10^8 \, \si{M_\odot}$, i.e. 100 times the DM mass of \vsmdpl. 
In the second step, for the MPBs that are not terminated before $z=5.89$, we used \florahg to further generate their progenitors.
Note that currently, we do not carry the hidden states between models (i.e. from \florahv to \florahg).
This is because the two models are trained independently due to the lack of N-body simulations spanning the entire redshift range from $z=0$ to $z \approx 24$, and hence their hidden states are not necessarily correlated. 
Thus here, for the sake of simplicity, we initialize the hidden state for \florahg with zeros but plan on investigating better initialization procedures in the future.
We chose a list of \gureft scale factors by sampling every 2 -- 6 \gureft snapshots from $z=5.89$ up to $z_\mathrm{max, gen}=24$.
Again, we set $z_\mathrm{max, gen}=24 > z_\mathrm{max, train}$ of \gureftc to demonstrate the model's extrapolation capability. 
Unlike in \vsmdpl, the $M_\mathrm{min, halo}$ is set based on the progenitor mass at $z=5.89$.
After we generate the full MPB, we apply the post-processing steps in Section~\ref{section:generation}.

We generated $387,031$ MPBs, the same number of trees as the \vsmdpl test dataset.
For convenience, we denote these combined MPBs as \florahvg MPBs. 

\medskip

We first present the median MAHs of \vsmdpl, \gureftc, and \florahvg from $z=0 - 24$ in Figure~\ref{fig:mean_mah_sim_florah}.
The median MAHs of \vsmdpl and \gureftc are represented by the data points, while the median MAHs of \florahvg are represented by the dashed lines.
The vertical dashed line at $z \approx 6$ represents the transition redshift at which we switch from \florahv to \florahg during the generation procedure. 
Overall, the \florahvg MAHs lines up nicely with both \vsmdpl at low redshifts and \gureftc at high redshifts. 
However, as mentioned in Section~\ref{section:simulation}, due to the resolution limit of \vsmdpl, less massive root halos of \vsmdpl ($\lesssim 10^{10} \si{M_\odot}$) do not have MAHs that overlap with \gureftc.
We cannot extend the MPBs of these halos with \gureftc and \florahg unless we allow the generation of poorly resolved halos (i.e. halos with fewer than 100 DM particles) during the first step of the generation procedure.
Thus, for halos with $\lesssim 10^{10} \si{M_\odot}$, which corresponds to the host halos of dwarf galaxies, we are unable to reliably generate their assembly histories beyond $z \approx 4 - 5$.
The longest \florahvg MPBs have root masses from $1.8 \times 10^{10} - 10^{11} \, \si{M_\odot}$ (corresponding to the bright end of the dwarf galaxy population) extend to $z \approx 20-22$.
Finally, it is worth noting that \gureft-90 and \vsmdpl do not overlap due to the mass cut applied when combining the \gureft boxes (as described in Section~\ref{section:combine_box}).
As a result, we do not fully utilize \gureft-90 during the generation. 
In future work, we can connect the massive end of \gureft-90 with \vsmdpl by including an additional simulation in the training dataset.

Next, we examine the generated MPBs in more detail.
Since we do not have N-body MPBs that span from $z=0$ to $z \approx 24$, we divide each \florahvg MPB into a low-redshift component which spans $z=0-10$ and a high-redshift component which spans $z=5.89-25$. 
We will compare the low-redshift component with \vsmdpl and the high-redshift component with \gureft.

\begin{figure*}
    \centering
    \includegraphics[width=\linewidth]{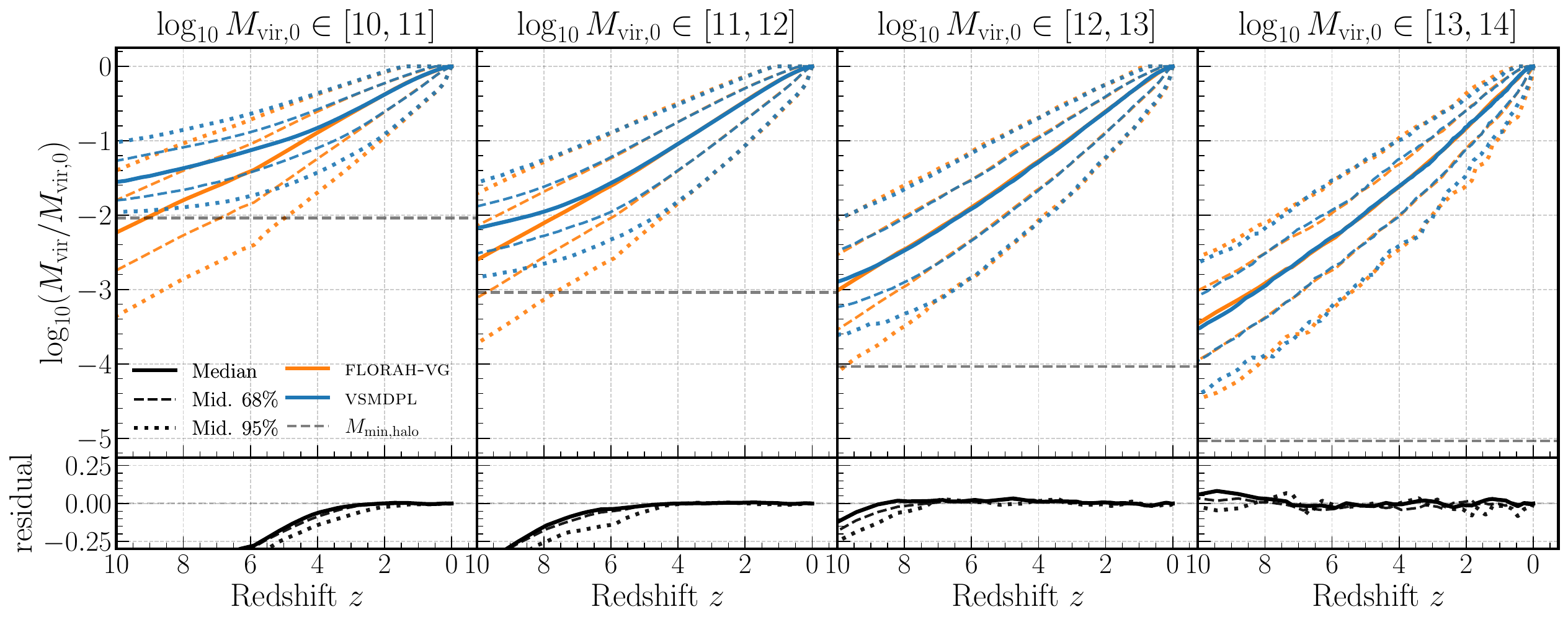}
    \includegraphics[width=\linewidth]{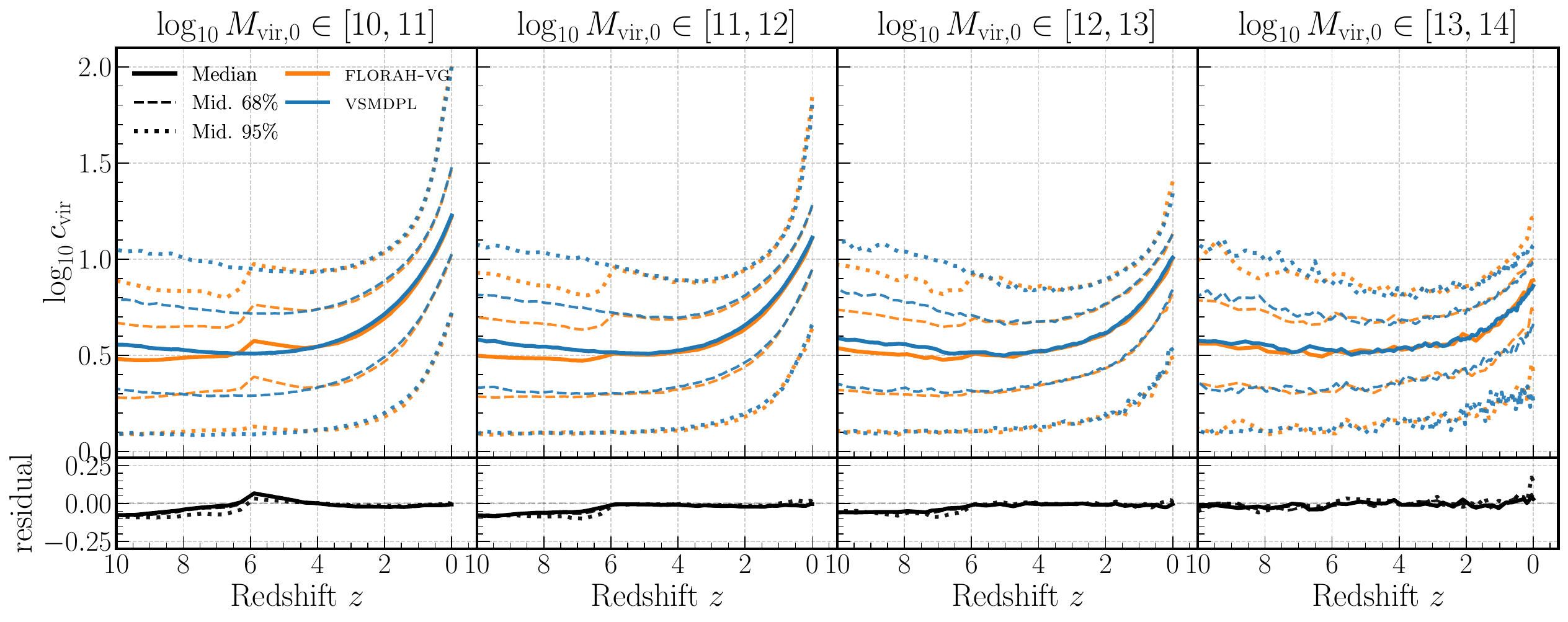}
    \caption{
    The MAHs and DM concentration histories of the MPBs in \vsmdpl and \florahvg in four mass bins (in \si{M_\odot unit}). 
    Panels are the same as Figure~\ref{fig:vsmdpl_mah_cah}.
    The disagreement between the \vsmdpl MAHs and \florahvg MAHs at high redshift is expected.
    Similarly to Figure~\ref{fig:vsmdpl_mah_cah}, the \vsmdpl MAHs plateau out at the resolution limit of \vsmdpl.
    On the other hand, at $z \gtrsim 6$, \florahvg MAHs do not plateau out since they are generated by \florahg, which is trained on \gureftc, and thus have higher resolutions than \vsmdpl. 
    }
    \label{fig:vsmdpl_gureft_lowz_mah_cah}
\end{figure*}

\begin{figure*}
    \centering
    \includegraphics[width=\linewidth]{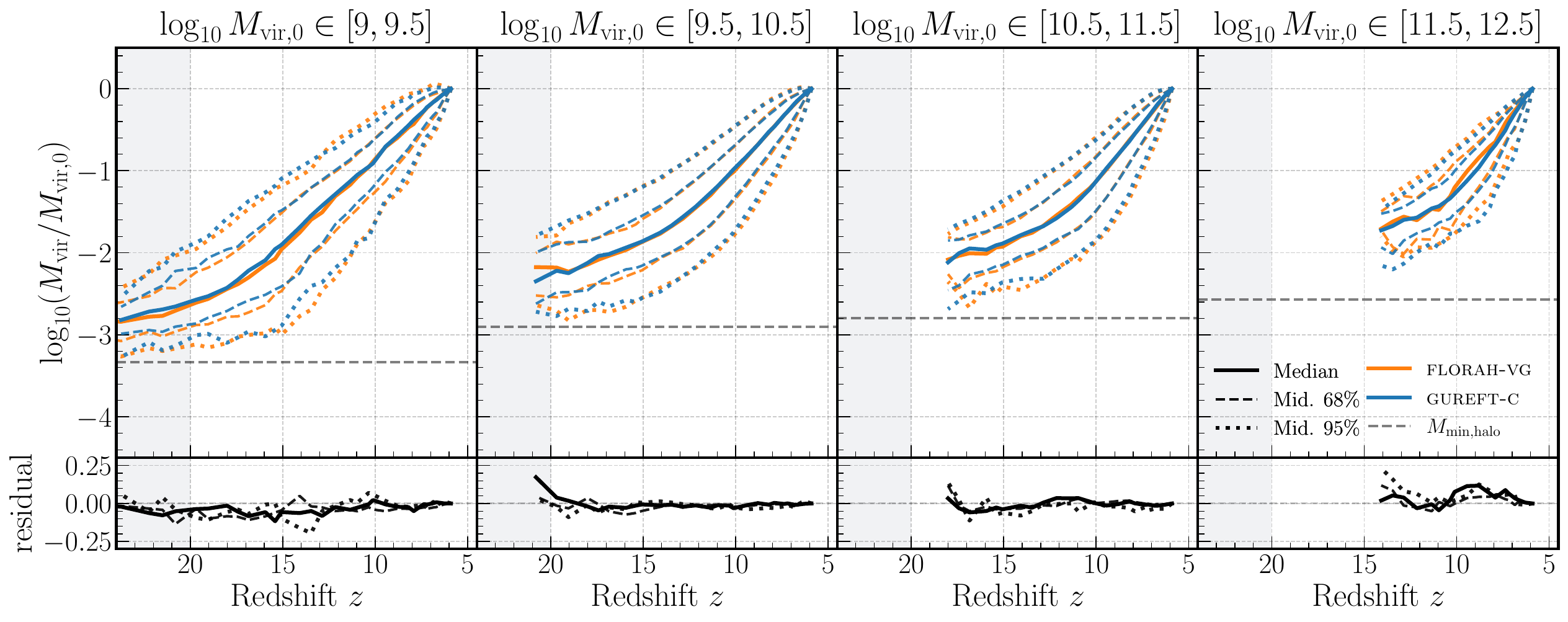}
    \includegraphics[width=\linewidth]{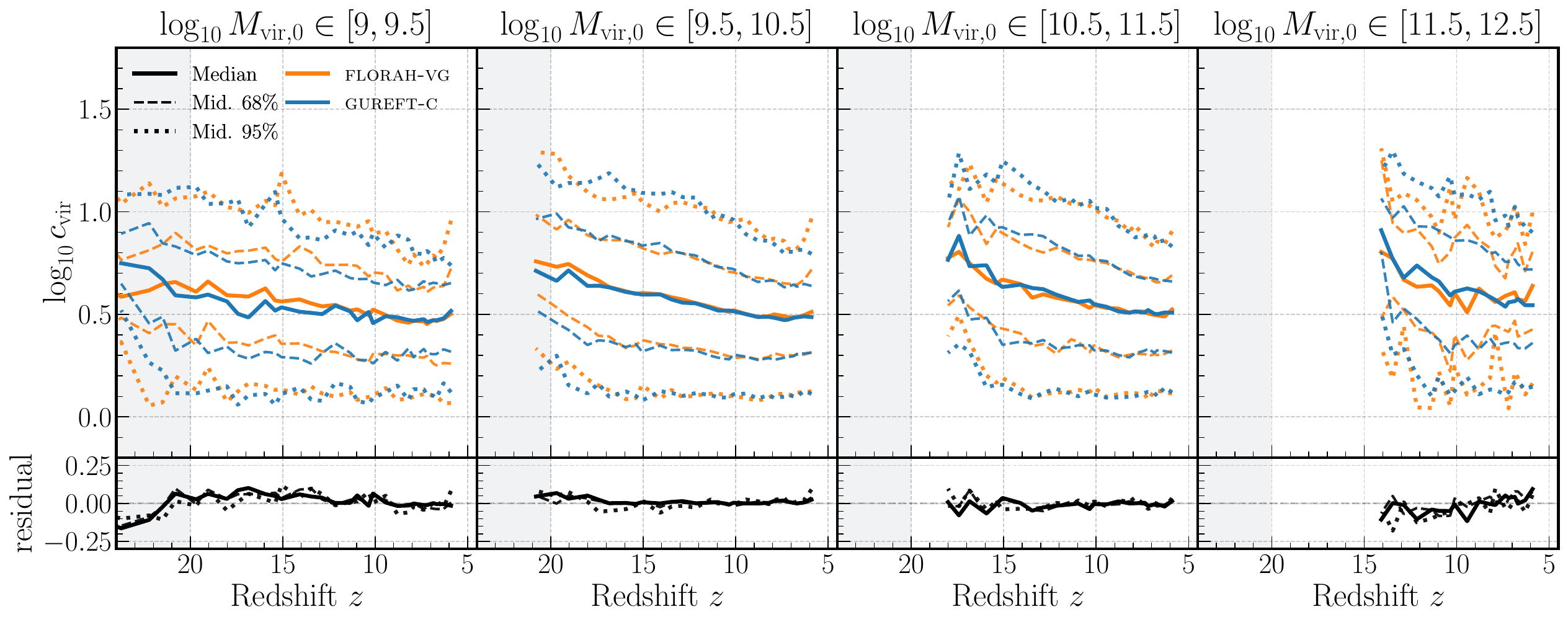}
    \caption{
    The MAHs and DM concentration histories of the MPBs in \gureftc and \florahvg in four mass bins (in \si{M_\odot unit}). Panels are the same as Figure~\ref{fig:vsmdpl_mah_cah}.
    }
    \label{fig:vsmdpl_gureft_highz_mah_cah}
\end{figure*}

\subsubsection{Mass and DM concentration histories at low redshifts}
\label{section:result_vsmdpl_gureft_lowz}

In Figure~\ref{fig:vsmdpl_gureft_lowz_mah_cah}, we show the median, 68\%-percentile, and 95\%-percentile containment regions of the MAHs (top panels) and DM concentration (bottom panels) histories for MPBs from \vsmdpl (dashed blue) and \florahvg (solid orange).
The MAHs and DM concentration histories of \vsmdpl are the same as Figure~\ref{fig:vsmdpl_mah_cah}.
Unlike the \vsmdpl MAHs and the \florahg MAHs in Figure~\ref{fig:vsmdpl_mah_cah}, the \florahvg MAHs do not plateau out at the resolution limit of \vsmdpl (horizontal dashed line).
This is to be expected because the progenitors of \florahvg halos at high redshifts are generated by the \florahg model and thus have a higher mass resolution.
We will show in Section~\ref{section:result_vsmdpl_gureft_highz} below that the MAHs at high redshifts are indeed consistent with \gureftc. 
However, we note some unphysical features in the MAHs and concentration histories.
The MAHs of \vsmdpl and \gureftc do not line up perfectly at the transition redshift $z=5.89$, creating ``kinks'' in the 95\% of the MAHs in the first two mass bins. 
This is particularly clear in the bottom panels, where the kinks can also be seen in the third mass bins and the median and 68\% of the DM concentration histories.
We defer the discussions of the limitations of our approach in more detail in Section~\ref{section:discussion}, after we compare the high-redshift components of \florahvg to \gureftc.

\subsubsection{Mass and DM concentration histories at high redshifts}
\label{section:result_vsmdpl_gureft_highz}

In Figure~\ref{fig:vsmdpl_gureft_highz_mah_cah}, we compare the high-redshift components of the \florahvg MPBs with \gureftc MPBs. 
As above, the median, 68\%-percentile, and 95\%-percentile containment regions of the MAHs (top panels) and DM concentration (bottom panels) histories.
Because \gureftc contains significantly fewer MPBs than generated by \florahvg (since \florahvg MPBs are generated from the root halos of \vsmdpl), we subsample \florahvg MPBs to match the number of MPBs in \gureftc for a fair comparison.
Similarly to Figure~\ref{fig:gureft_mah_cah}, we do not include redshift slices with fewer than 10 resolved halos, resulting in the MAHs and concentration histories in the last two bins terminating before $z_\mathrm{max, train}=20$. 
In general, both the MAHs and concentration histories of \florahvg agree well with \gureftc. 
Unlike in the low-redshift case, the MAHs of both \gureftc and \florahvg plateau out at $M_\mathrm{min, halo}$ as expected.
As a reminder, here $M_\mathrm{min, halo}$ is determined by the $\log_{10} M_\mathrm{vir, 0}$ at $z=5.89$ (refer Table~\ref{tab:sim}).
Note that we still see the effect of the sharp transition between \vsmdpl and \gureftc at $z=5.89$ in the concentration histories, though to a much lesser extent.

\section{Discussion}
\label{section:discussion}

\subsection{Comparison with previous works}
Previous approaches to modeling the cosmological MAHs of DM halos typically fit a parameterized functional form over the entire simulated MAH population (e.g.~\citealt{2021OJAp....4E...7H, 2009MNRAS.398.1858M, 2001astro.ph.11069W}, etc.).
For example, \cite{2009MNRAS.398.1858M} assumes the MAHs take the form:
\begin{equation}
    M(z) = M_0 (1 + z)^\beta e^{-\gamma z},
\end{equation}
where $M_0 = M(0)$ is the mass of the root halo, and $\theta \equiv (\beta, \gamma)$ are free parameters of the fit. 
One can also imagine fitting the evolution of the DM concentration $\cvir(z)$ or other properties of the halos in a similar manner.
As shown in the aforementioned works, these approaches can also provide a reasonably accurate representation of the MAH and capture population-level statistics such as the mean MAH for a given root halo mass. 
To capture the full distribution (i.e., the scatter) of the MAH population, one can model the full distribution of the fit parameters $P(\theta)$, as shown in \cite{2021OJAp....4E...7H}.
The dependency of $\theta$ on the root halo properties such as $M_0$ and $c_0$ can be included as a conditional probability, i.e. $P(\theta | M_0, c_0)$.

The \florah framework can be considered as a non-parametric modeling method for the MAH, with a few notable advantages.
First, the parametric modeling approaches above assume a functional form not only for the mean of the MAH but also the scatter (and other quantiles) of the MAH population distribution, as this stochasticity is folded into the sampling of the fit parameters $P(\theta | M_0, c_0)$ at $z=0$.
It is unclear if these all should follow the same distribution.
Thus, non-parametric methods such as \florah can provide more flexibility in modeling the MAH distribution.
Moreover, parametric methods may require additional assumptions to incorporate more than one halo property.
For example, to fit both the MAH and the DM concentration histories, one needs to assume how the mass and concentration are correlated, e.g. via the halo concentration-mass relations.
On the other hand, \florah can jointly fit both the mass and DM concentration \textit{without requiring additional assumptions about the data}.
Indeed, we show in Figures~\ref{fig:vsmdpl_cvir_mass} and \ref{fig:gureft_cvir_mass} that MPBs generated by \florah follow the same halo concentration-mass relations as the simulations.
Additionally, incorporating additional halo properties (e.g., the halo shape or environment information) into \florah is straightforward, since the conditional normalizing flows can easily model high-dimensional distributions. 
We plan to explore different sets of halo properties in future work.

As shown in Section~\ref{section:result}, \florah can correctly capture the progenitor-descendant mass ratios at each time step and model the MAH at the individual halo level.
The distribution of the mass (and DM concentration) at each redshift $z$ is modeled independently with the conditional normalizing flows, thus allowing for ``jumps'' in the MAH to occur, representing DM halo mergers, which can play an important role in galaxy formation models.
In contrast, the parametric modeling approaches assume the MAH to be a smooth function, so there is no stochasticity between redshifts once the fit parameters $\theta$ are sampled from $P(\theta | M_0, c_0)$ at $z=0$.
Though it is possible to increase the fidelity of the parametric fits by resampling $\theta$ at different $z$, this requires modeling $P(\theta | M_0, c_0, z)$, which will be much more difficult.
Halo-level properties such as the progenitor-descendant mass ratios are important for certain modeling applications of black hole seeding in SC-SAM, which we will explore with \florah in future works.
We note that EPS-based approaches can also model the stochasticity between redshifts in a similar autoregressive manner, by sampling the conditional mass function from the EPS formalism. 
However, in EPS-based approaches, the MAH is usually modeled as a Markov Chain, in which the mass of the progenitor is modeled using only the mass of its immediate descendant. 
\florah extends this so that the progenitor mass is modeled using the entire assembly history.

\subsection{Current limitations and future outlook}

We showed in Section~\ref{section:vsmdpl_result} that when trained on a single N-body simulation, \florah can accurately recover the MAHs and concentration histories. 
This applies similarly when multiple N-body simulations with the same redshift range but varying mass resolutions are combined into a single training dataset (Section~\ref{section:gureft_result}).
We further showed that \florah-generated MAHs can be input into the SC-SAM to predict observable properties and assembly bias of galaxies.
However, because the current framework only generates the MPBs of merger trees, \florah cannot yet capture correctly merger-driven properties such as the stellar mass of some population of galaxies or the mass of supermassive black holes. 
To capture these properties, we must include secondary branches as well as other sub-branches in the generation process.
In ongoing work, we are extending it to generate full merger trees by, for example, modeling the probabilities of branching events and utilizing graph generative models.

In Section~\ref{section:gureft_vsmdpl_result}, we combined the \florahv and \florahg model to generate assembly histories from $z=0$ to $z \approx 24$.
Overall, we found that the generated MAHs and concentration histories agree with the N-body simulations. 
However, the current approach has two main drawbacks.

First, due to the lack of N-body simulations spanning the redshift range from $z=0$ to $z \approx 24$, \florahv and \florahg are trained independently.
Thus, their hidden states are not necessarily correlated and cannot be carried from one model to the other (i.e. from \florahv to \florahg). 
As a result, the information carried by the hidden states of \florahv is lost during the transition.
We plan on investigating better initialization procedures in future work.
For example, instead of transitioning at $z = 5.89$, we may better utilize the overlapping redshift range between \vsmdpl and \gureft.
We found that the current range, $z > 5.89$, to be insufficient because it includes too many poorly resolved halos of \vsmdpl (i.e. those with fewer than 100 DM particles), which create an inconsistency between the MAHs of \vsmdpl and of \gureft.

Second, the current approach for combining the simulations can result in some unphysical features in the MAHs and concentration histories at the transition redshift $z=5.89$ between the two simulations. 
This is especially clear in the concentration histories, where we see that the concentration histories of \vsmdpl are consistently higher than \florahvg at $z > 5.89$.
However, from Figure~\ref{fig:vsmdpl_gureft_highz_mah_cah}, it is clear that the concentration histories of \florahvg is consistent with \gureftc.
This implies that there is a systematic difference between the concentration histories of \gureftc and \vsmdpl. 
We attribute this difference to the absence of unresolved, low-mass progenitor halos in \vsmdpl at these redshifts.
At high redshifts, \cvir tends to be lower for lower-mass halos, so missing the low-mass halo population will artificially push the distribution of \cvir towards the high values.
This is supported by the fact that the lower-68\% and lower-95\% of the concentration histories are less affected compared to the upper-68\% and upper-95\%.
Although a more sophisticated generation procedure may mitigate the sharp transitions, we attribute the problem mainly to the inconsistency in the MAHs and concentration histories of the two training simulations. 
We refer to \cite{gureft_paper} for a more detailed diagnosis of this inconsistency and comparison between \gureft and the MultiDark simulations (including \vsmdpl).

Both of the above drawbacks can be mitigated by improving the simulations.
In future work, we are considering the possibility of further running \gureft to a redshift of 4 and with a larger box size (e.g. 120 \si{Mpc \, h^{-1}}).
This will extend the overlapping redshift range between the two simulations and ensure that \vsmdpl has a sufficient number of resolved halos in this range, thus improving the consistency between the assembly histories.
In addition, this enables the assembly histories of halos with less than $10^{10} \, \si{M_\odot}$ to be extended using a \gureft-based \florah model.

\section{Conclusion}
\label{section:conclusion}

In this paper, we developed \florah, a deep generative model based on recurrent neural networks and normalizing flows, to generate the mass assembly histories (MAHs) of dark matter halo merger trees.
We trained two models independently on the \vsmdpl and \gureft N-body simulations and assessed the performance of the models to recover the MAHs, concentration histories, and observable and semi-observable properties derived from semi-analytic models such as stellar mass. 
Our main findings are summarized as follows:
\begin{itemize}
    \item In Section~\ref{section:vsmdpl_result}, we found that the \florahv model, which is trained on the \vsmdpl simulation up to redshift 10, can accurately capture the MAHs and concentration histories across a wide range of host halo masses.
    Additionally, when operating well above the resolution limit of the simulation,  \florahv robustly extrapolates the histories up to redshift 14, beyond the training redshift.
    We applied the Santa Cruz semi-analytic model to \florah-generated histories and showed that \florah can recover correlations between galaxy properties and halo assembly history, such as the correlation between stellar-to-mass ratio residual and halo concentration, as well as the galaxy matter power spectrum. 
    \florah is the first machine learning framework that accurately learns and incorporates the correlations between halo properties and merger history.

    \item  In Section~\ref{section:combine_box}, we developed a procedure to incorporate multiple simulations with the same redshift range and varying mass resolutions into a single training dataset.
    In Section~\ref{section:gureft_result}, we trained the \florahg model on the \gureft suite of simulations, comprising four boxes (\gureft-05, \gureft-15, \gureft-35, and \gureft-90), and successfully demonstrated \florah's ability to handle this variation. 
    The \gureft simulations were purposefully crafted to capture the evolution of dark matter halos at the ultra-high redshift Universe and at an unprecedented temporal resolution.
    By successfully learning the assembly histories of these simulations, \florah proves to be a versatile tool for studying the formation and evolution of cosmic structures during the initial phases of the Universe.

    \item In Section~\ref{section:gureft_vsmdpl_result}, we developed a procedure to concatenate multiple \florah models.
    We used \florahv to generate the assembly histories from $z=0$ to $z=5.89$ and \florahg to further generate the histories to $z\approx20-24$.
    We showed that despite the simplicity of our approach, we were able to generate the mass assembly histories from $z=0$ to an ultra-high redshift $z \approx 24$ for halos with root mass $\lesssim~10^{10} \, \si{M_\odot}$, which corresponds to the bright end of the dwarf galaxy population. 
    We were unable to reliably generate assembly histories below this mass due to the resolution limitation of \vsmdpl.
    In addition, during the generation process, we found an inconsistency in the dark matter concentration histories of \vsmdpl and \gureft at $z \gtrsim 6$, which may be attributed to the missing of unresolved progenitor halos in \vsmdpl (refer to \cite{gureft_paper} for a more detailed discussion). 
    This results in sharp kinks in the concentration histories at the transition redshift $z = 5.89$ between the two \florah models.
    These limitations can be overcome by improving the overlapping between \gureftc and \vsmdpl, e.g., by running \gureftc to a lower redshift and with a larger volume.)
    Despite the current limitations, this represents the first step towards generating assembly histories at a mass resolution required to simultaneously capture the merger dynamics from dwarf galaxies to galaxy clusters up to high redshifts and in large volumes.
    At present, this is far beyond the capability of numerical simulations.

    \item In Section~\ref{section:discussion}, we discuss the advantages of \florah over traditional parametric modeling approaches for the MAHs.
    Like many machine learning methods, \florah is non-parametric and does not require assumptions on the functional form of the mass and DM concentration assembly histories.
    It is also straightforward to include additional halo properties beyond the mass and DM concentration since the normalizing flows employed for density modeling excel at capturing high-dimensional distributions. 
    In addition, \florah models the MAHs in an autoregressive manner, sampling the halo mass and concentration at each redshift based on the entire halo history. 
    This allows \florah to capture variations between redshifts, such as the progenitor-descendant mass ratios, more easily. 
    
\end{itemize}

\florah represents an exciting and promising initial step towards the development of a machine learning-based framework for generating full merger trees. 
Such a framework has the potential to revolutionize our understanding of how galaxies form and evolve by allowing for the exploration of different galaxy formation scenarios with excellent computational efficiency at unprecedented accuracy.
To achieve this goal, we intend to expand \florah's capabilities to generate secondary branches and other sub-branches. 
In addition, we aim to develop an emulator that can generate merger trees by training \florah on simulations featuring different cosmological parameters. 
Doing so will enable \florah to learn and capture the dependence of halo assembly histories on cosmology.

\section*{Software}
This research made use of the 
\texttt{corner}~\citep{corner},
\texttt{IPython}~\citep{PER-GRA:2007},
\texttt{Jupyter}~\citep{Kluyver2016JupyterN},
\texttt{Matplotlib}~\citep{Hunter:2007},
\texttt{NumPy}~\citep{harris2020array},
\texttt{nflows}~\citep{nflows},
\texttt{PyTorch}~\citep{NEURIPS2019_9015}, 
\texttt{PyTorch Lightning}~\citep{william_falcon_2020_3828935},
\texttt{SciPy}~\citep{2020NatMe..17..261V},
and \texttt{ytree}~\citep{ytree}
software packages.

\section*{Data Availability}

The code version used for this article is available at~\url{https://github.com/trivnguyen/florah}. 
The \vsmdpl training dataset in this article uses public simulations from the CosmoSim and MultiDark database available at~\url{https://www.cosmosim.org/}.
The \gureft training dataset will be shared on reasonable request to the corresponding author. 
The output data generated by the \florah models will be shared on reasonable request to the corresponding author.

\section*{Acknowledgements}

We thank Christian Kragh Jespersen and Yuan-Sen Ting for meaningful discussions. 

TN, CM, and RS are supported by the Center for Computational Astrophysics at the Flatiron Institute.
The Center for Computational Astrophysics at the Flatiron Institute is supported by the Simons Foundation.
The computations in this work were, in part, run at facilities supported by the Scientific Computing Core at the Flatiron Institute, a division of the Simons Foundation.
The data used in this work were, in part, hosted on equipment supported by the Scientific Computing Core at the Flatiron Institute, a division of the Simons Foundation.
TN is also supported by the National Science Foundation under Cooperative Agreement PHY-2019786 (The
NSF AI Institute for Artificial Intelligence and Fundamental Interactions, http://iaifi.org/).
AY is supported by an appointment to the NASA Postdoctoral Program (NPP) at NASA Goddard Space Flight Center, administered by Oak Ridge Associated Universities under contract with NASA.

The CosmoSim database used in this paper is a service by the Leibniz-Institute for Astrophysics Potsdam (AIP).
The MultiDark database was developed in cooperation with the Spanish MultiDark Consolider Project CSD2009-00064.
The authors gratefully acknowledge the Gauss Centre for Supercomputing e.V. (www.gauss-centre.eu) and the Partnership for Advanced Supercomputing in Europe (PRACE, www.prace-ri.eu) for funding the MultiDark simulation project by providing computing time on the GCS Supercomputer SuperMUC at Leibniz Supercomputing Centre (LRZ, www.lrz.de).
The Bolshoi simulations have been performed within the Bolshoi project of the University of California High-Performance AstroComputing Center (UC-HiPACC) and were run at the NASA Ames Research Center.



\bibliographystyle{mnras}
\bibliography{florah} % if your bibtex file is called example.bib

%\begin{thebibliography}{99}
%\end{thebibliography}

%%%%%%%%%%%%%%%%%%%%%%%%%%%%%%%%%%%%%%%%%%%%%%%%%%

%%%%%%%%%%%%%%%%% APPENDICES %%%%%%%%%%%%%%%%%%%%%

% \appendix
% \input{appendix}

%%%%%%%%%%%%%%%%%%%%%%%%%%%%%%%%%%%%%%%%%%%%%%%%%%

% Don't change these lines
\bsp	% typesetting comment
\label{lastpage}
\end{document}

%% file: table_sim.tex
\begin{table*}
\caption{
The simulation specifications of \vsmdpl and the four \gureft simulations. The third and second-to-last columns show the minimum and maximum root mass of the training dataset for each simulation. Note that the four \gureft simulations are combined into one training dataset. The last column shows the minimum progenitor mass during the generation process.
}
\label{tab:sim}
\begin{tabular}{ccccccccccc}
\hline
& Box size          & $M_\mathrm{DM}$                   & N        & $\epsilon$ & $n_\mathrm{snap}$ & Redshift range & min $M_\mathrm{vir, 0}$   & max $M_\mathrm{vir, 0}$ & $M_\mathrm{min, halo}$\\
& [\si{Mpc \, h^{-1}}] & [\si{\mathrm{M}_\odot \, h^{-1}}] &          & [\si{kpc \, h^{-1}}]  & & &  [\si{\mathrm{M}_\odot \, h^{-1}}] &  [\si{\mathrm{M}_\odot \, h^{-1}}] & [\si{\mathrm{M}_\odot \, h^{-1}}]\\ \hline
\gureft-05 & $5$               & $9.92 \times 10^3$                & $1024^3$ & $0.16$   &  171  &  $5.89-40.0$  & $4.96 \times 10^6$ & $2.14 \times 10^9$ &  $9.92 \times 10^5$ \\
\gureft-15 & $15$              & $2.68 \times 10^5$                & $1024^3$ & $0.49$   &  171  &  $5.89-40.0$  & $2.14 \times 10^9$ & $2.14 \times 10^{10}$ &  $2.68 \times 10^7$   \\
\gureft-35 & $35$              & $3.40 \times  10^6$               & $1024^3$ & $1.14$   &  171  &  $5.89-40.0$  & $2.14 \times 10^{10}$ & $2.14 \times 10^{11}$ &   $3.40 \times 10^8$ \\
\gureft-90 & $90$              & $5.78 \times 10^7$                & $1024^3$ & $2.93$   &  171  &  $5.89-40.0$  & $2.14 \times 10^{11}$ & n/a &   $5.78 \times 10^9$  \\ \hline
\vsmdpl    & $160$             & $6.20 \times 10^6$                & $3840^3$ & $1.0 \rightarrow 2.0$ & 151 & $0.00-24.9$ & $3.10 \times 10^8$ & n/a & $6.20 \times 10^8$ \\ \hline 
\end{tabular}
\end{table*}